\documentclass[12pt,preprint]{aastex}
\usepackage{graphics,graphicx}
\begin{document}

\title{X-ray Emission from Orion Nebula Cluster Stars with
  Circumstellar Disks and Jets}

\author{Joel H. Kastner\altaffilmark{1}, Geoffrey
  Franz\altaffilmark{1}, Nicolas Grosso\altaffilmark{2},
  John Bally\altaffilmark{3}, Mark J. McCaughrean\altaffilmark{4},
  Konstantin Getman\altaffilmark{5}, Eric
  D. Feigelson\altaffilmark{5}, Norbert  
  S. Schulz\altaffilmark{6}}

\altaffiltext{1}{Chester F. Carlson Center for Imaging
Science, Rochester Institute of Technology, 54 Lomb Memorial
Dr., Rochester, NY 14623; jhk@cis.rit.edu}
\altaffiltext{2}{Laboratoire d'Astrophysique de Grenoble,
  Universite Joseph-Fourier, 38041 Grenoble Cedex 9, France} 
\altaffiltext{3}{Center for Astrophysics and Space
  Astronomy, University of Colorado, 389 UCB, Boulder, CO
  80309-0389} 
\altaffiltext{4}{University of Exeter, School of Physics,
  Stocker Road, Exeter EX4 4QL, Devon , UK; and
  Astrophysikalisches Institut Potsdam, An 
  der Sternwarte 16, 14482 Potsdam, Germany} 
\altaffiltext{5}{Department of Astronomy and Astrophysics,
  Pennsylvania State University, 525 Davey 
  Laboratory, University Park, PA 16802} 
\altaffiltext{6}{Center for Space Research, MIT, Cambridge, MA 02139}

\begin{abstract}
We investigate the X-ray and near-infrared emission
properties of a sample of pre-main sequence (PMS) stellar
systems in the Orion Nebula Cluster (ONC) that display evidence
for circumstellar disks (``proplyds'') and optical jets in 
Hubble Space Telescope (HST) imaging. Our study uses X-ray  
data acquired during Chandra Orion Ultradeep
Program (COUP) observations, as well as complementary optical and
near-infrared data recently acquired with HST and the Very
Large Telescope (VLT), 
respectively. Approximately 70\% of $\sim$140 proplyds 
were detected as X-ray sources in
the 838 ks COUP observation of the ONC, including $\sim25$\%
of proplyds that do not display central stars in HST
imaging. In near-infrared imaging, the detection rate of proplyd
central stars is $>90$\%. Many proplyds display near-infrared 
excesses, suggesting disk accretion is ongoing onto the
central, PMS stars. About 50\% of circumstellar disks that are detected
in absorption in HST imaging 
contain X-ray sources. For these sources, we find that X-ray
absorbing column and apparent disk inclination are well
correlated, providing insight into the
disk scale heights and metal abundances of UV- and X-ray-irradiated
protoplanetary disks. 

Approximately 2/3 of the $\sim30$ proplyds and PMS stars
exhibiting jets in Hubble images have COUP X-ray
counterparts. These jet sources display some of the largest
near-infrared excesses among the proplyds, suggesting that
the origin of the jets is closely related to ongoing, PMS
stellar accretion. One morphologically complex jet source,
d181--825, displays a double-peaked X-ray spectral energy
distribution with a prominent soft component that is
indicative of strong shocks in the jet collimation region. A
handful of similar objects also display X-ray spectra that
are suggestive of shocks near the jet source. These results
support models in which circumstellar disks collimate and/or
launch jets from young stellar objects and, furthermore,
demonstrate that star-disk-jet interactions may contribute
to PMS X-ray emission.
\end{abstract}


\section{Introduction}

Some of the best examples of circumstellar disks
around low-mass, pre-main sequence (PMS) stars
are found among the members of the \object{Orion Nebula
Cluster} (ONC). These objects, detected in {\it Hubble Space
Telescope} (HST) imaging and dubbed ``proplyds'' (short for
``protoplanetary disks'') by O'Dell and collaborators
(O'Dell \& Wong 1996 and references therein), are seen in
projection in front of (or lie embedded within) the Orion
Nebula. The morphologies of proplyds\footnote{The term
proplyd is used throughout this paper to refer to apparent PMS
circumstellar disk systems detected in HST imaging, but use
of this term is not intended to suggest that the status of
such systems as
planet formation sites is well established.} seen in HST
imaging range from cometary globules that are externally
illuminated and/or ionized to structures resembling dusty
disks seen in silhouette against the bright nebular
background\footnote{Similar silhouette structures were noted
by Feibelman (1989), based on examinations of deep
photographs of the Orion Nebula.} (McCaughrean \& O'Dell
1996; Bally, O'Dell, \& McCaughrean 2000). In addition, many
proplyds and other ONC members are observed to drive
collimated jets. Such jets are detected on scales ranging
from the subarcsecond (``microjets''; Bally et al.) to many
arcminutes (Smith et al.\ 2005; Bally et al.\ 2005) as a
consequence of the emission from ionized gas close to the
central stars and of the large-scale chains of knots of
shock-excited emission (known as Herbig-Haro [HH] objects)
powered by the outflowing gas, respectively. 

The origins of PMS jets are, presumably, intimately related to the
presence of circumstellar disks, as present theory holds
that these disks provide the jet launching and/or
collimation mechanisms (as first proposed by Blandford \&
Payne [1982] in the context of black hole accretion
disks). Several such mechanisms have been proposed to
explain PMS jets and outflows, with most of these mechanisms
invoking disk and/or stellar magnetic fields (e.g., the
so-called ``X-wind'' model, Shu et al.\ 1988, 1995; see also
Goodson, Winglee, \& Boehm 1997; Turner, Bodenheimer, \& Rozyczka 1999;
Delamarter, Frank, \& Hartmann 2000; Matt et al.\
2003; a nonmagnetic outflow launching model was proposed by Soker \& Regev
2003). Hence, ONC sources exhibiting microjets offer a probe 
of the disk-jet connection in low-mass, PMS stars.

A key result of the first {\it Chandra X-ray Observatory}
(CXO) observations of the ONC was the detection of a number
of X-ray sources at or near the positions of proplyds
(Garmire et al.\ 2000; Schulz et al.\ 2000). While proplyd
X-rays are most readily attributed to magnetic activity
associated with their host PMS stars (Feigelson \&
Montmerle 1999; Favata \& Micela 2003), plasma production
from magnetic star-disk-jet interactions may also play a
role. In this respect, the detection of X-ray emission from
proplyds, and further characterization of their X-ray
emission properties, should inform the present debate
concerning the processes responsible for X-ray emission from
low-mass, PMS stars in general (see, e.g., discussions in
Kastner et al.\ 2004 and Preibisch et al.\
2005). Furthermore, the attenuation of proplyd X-ray sources
by circumstellar material can be used as a unique probe of
the density structure of protoplanetary disks and of X-ray
irradiation of circumstellar disks by the central T Tauri
stars.

Soft and/or diffuse X-ray emission has also been
detected in association with several protostellar outflows 
(e.g., Pravdo et al.\ 2001, 2004; Favata et al.\ 2002; Bally
et al.\ 2003; Tsujimoto et al. 2004). Such emission 
presumably originates from energetic shocks generated by
collisions between protostellar jets and ambient
molecular cloud material (e.g., Pravdo et al.\ 2001; Bonito
et al.\ 2004). Despite the frequent association of both jets
and X-rays with proplyds, however, it has yet to be
established whether any proplyds actually produce
X-ray emission via shocks. 

On a more fundamental level, the nature of proplyds
that lack central stars in HST imaging (Bally et al.\ 2000)
remains uncertain. Although these objects resemble
morphologically those proplyds that clearly consist of
envelopes and/or disks surrounding low-mass, PMS stars,
the PMS evolutionary status of ``starless'' proplyds has
yet to be firmly established. High-resolution X-ray
and near-infrared imaging provides an excellent means to determine
whether, in fact, such objects contain central, PMS stars.

The \dataset[ADS/Sa.CXO#defset/COUP]{Chandra Orion Ultradeep
  Program} (COUP) observation of
the ONC (Getman et al.\ 2005a) has resulted in the detection
of $\sim1400$ X-ray emitting PMS stars in the ONC (Getman et al.\ 2005b), 
including the majority of previously catalogued proplyds. In
this paper, we investigate the X-ray emission 
and optical/infrared properties of these objects. Images recently
acquired with the Advanced Camera for Surveys (ACS) aboard
the Hubble Space Telescope (HST) provide improved optical
positions for proplyds. These images, in
combination with the COUP X-ray data as well as near-infrared
photometry acquired with the Very Large Telescope (VLT),
yield new insight into the nature of 
the ONC's proplyds and jet sources. 

In \S 2 the sample and data are described; \S 3 contains a
description of the results of the correlation of
X-ray source positions with optical (HST/ACS) and
near-infrared source positions; in \S 4,
results are presented for the X-ray counterparts to
circumstellar disks detected in absorption in the ACS
images and jet sources, and \S 5 contains results for
the X-ray and near-infrared emission
properties of ONC optical jet sources. A discussion of
these results is presented in \S 6. Sec.\ 7 contains a summary.

\section{The Proplyd Sample: HST \& COUP Observations}


The HST surveys of O'Dell \& Wong (1996) and Bally et al.\
(2000) constitute the basis for our 
identification of the COUP X-ray counterparts to ONC
proplyds. There is considerable overlap between these two 
surveys, with only four proplyds (159-418, d163-026s,
d172-028s, and d244-440) in Bally et al.\ not
included in the lists of O'Dell \& Wong (where the
proplyd names used here follow the 
nomenclature of O'Dell \& Wen [1994] and Bally et al.). Proplyds
140-512 and 141-520 in O'Dell \& Wong (1996) correspond to
the single proplyd d141-520 in Bally et al.\ (2000). After
removing such duplications, a total of 164 proplyds, proplyd
candidates, and jet (or wind) sources are identified by these
two surveys. To this 
total, we add 7 new proplyds identified in recent HST
Advanced Camera for Surveys (ACS) imaging by Smith et al.\
(2005), and 1 additional proplyd (044-527) identified
during the course of our own inspection of the ACS images
(\S 2.2). The complete sample of objects considered here is
listed in Tables 1 and 2, where Table 2 includes only those 
objects that do not appear as proplyds in ACS imaging (\S 2.2).

\subsection{Imaging with the HST Advanced Camera for
  Surveys}

The HST/ACS Wide Field Camera (WFC) images analyzed for this
paper constitute a subset of a mosaic, obtained during HST
Cycle 12, that covers more than 400 square arcmin. This
ASC Orion image set was obtained through the WFC's F658N
filter, such that the images include
H$\alpha$ $\lambda 6563$ and [N {\sc ii}] $\lambda 6583$
emission. The pixel scale was $0.05''$ pix$^{-1}$. The
images were calibrated astrometrically based on 2MASS
imaging of the ONC. These and other aspects of the
observations and data reduction are described in detail in
Bally et al.\ (2005; see also Smith et al.\ 2005). 

To verify the identifications of proplyds and jet sources,
and to ascertain the precise positions and morphologies of these
sources, we examined the ACS images at the
position of each of the 166 objects lying within the
field of view of the ACS mosaic. A description of the
appearance of each proplyd candidate is included in Tables 1
and 2. Generally, the objects listed in Table 1 fall into
one or both of two categories, as noted by Bally et al.\
(2000): (1) dark disks seen in absorption against the bright
nebular background, and (2) externally illuminated and/or
ionized globules (noted in Table 1 as ``cometary rim'' or
``cometary tail''). As noted in Table 1, some of the
proplyds also show evidence for collimated jets and/or
teardrop-shaped ionization fronts. The ionization fronts
result from photoablation and photoionization of
circumstellar material caused by the intense UV fields of the OB
stars in the Orion Trapezium cluster. 

For $\sim70$\% of the proplyds, central 
stars are apparent in the ACS images. The 
positions listed in Tables 1 and 2 are the positions of
these stars (in the case of the close binaries listed in
Table 2, the positions 
are those of the brighter component). For those objects with no
readily identifiable star, the position listed corresponds
to the apparent center of the dark lane within
and/or center of symmetry of the inner, circularly symmetric
portion of a cometary globule (see \S 3). We estimate --- and correlation
with COUP source positions (\S 3) confirms --- that the
positions listed in Tables 1 and 2 are typically accurate to
$\sim0.2''$. For proplyds with central stars detected
in ACS images, the positions are typically accurate to
$\sim0.1''$, i.e., similar to the astrometric uncertainties
of the 2MASS Point Source Catalog\footnote{See http://spider.ipac.caltech.edu/staff/hlm/2mass/overv/overv.html}.

We found no evidence for proplyd-like structures (such as just
described) at a number of positions previously ascribed to non-stellar
sources likely to be proplyds (O'Dell \& Wong 1996). Most of
these positions, which are listed in Table 2, 
correspond to apparently single stars, close binaries, or
Herbig-Haro objects (``HH 
knots''); in a few cases, there is no source readily
apparent at the listed position. We have 
eliminated these 22 objects from further consideration as
proplyds, although we do report their COUP counterparts (\S
3) in Table 2.

\subsection{COUP data}

The $\sim838$ ks COUP observation of the ONC represents the
richest source of X-ray data yet obtained for a young star
cluster. A complete description of the observations and of
the X-ray data reduction, source detection, spectral and
light curve extraction, and spectral fitting procedures is
contained in Getman et al.\ (2005a). The COUP observation resulted
in the detection of 1616 individual X-ray sources, with
typical formal positional uncertainties of $<0.3''$ (and often
$<0.1''$). The vast 
majority of these sources have been unambiguously identified with
pre-main sequence stars detected in the optical and/or
near-infrared;
overall, only $\sim10$\% of COUP sources are associated with
extragalactic objects, while $\sim1$\% are foreground stars
(Getman et al.\ 2005b). In the present 
paper, we make use of the association of
most COUP sources with near-infrared sources detected in
subarcsecond VLT imaging in the J (1.25 $\mu$m), H
(1.65 $\mu$m), and K (2.2 $\mu$m) bands (McCaughrean
et al., in prep.). The resulting photometry has been
converted to the 2MASS JHK$_s$ system, and merged with 2MASS
and other available near-infrared photometry to include sources
for which magnitudes cannot be 
obtained from the VLT images (due to detector saturation). 
Typical photometric uncertainties in the merged
near-infrared catalog are $<0.1$ mag. 
We also utilize the results of fits of one- or
two-component thermal plasma models to COUP X-ray spectra
(Getman et al.\ 2005a).
These fits yield estimates of the line-of-sight absorbing column
($N_{\rm H}$) to and the broadband ($0.5-8.0$ keV) X-ray luminosity of
each source. 

\section{X-ray and Near-infrared Counterparts to Proplyds}

We correlated the positions of COUP sources with the ACS
positions of all 172 potential proplyds in Tables
1 and 2. Tables 1--4 summarize the results of this ACS vs.\
COUP position correlation.  We find that all of the X-ray
counterparts to those proplyds in Table 1 that lie within $\sim2'$
of the {\it Chandra} boresight in the COUP image --- i.e.,
for which the {\it Chandra} ACIS-I image quality is
sufficient to determine whether or not a source is extended
with respect to the $<1.0''$ FWHM {\it Chandra} PSF,
without resorting to deconvolution techniques --- are
point-like.

From the initial set of 105 COUP source positions that lie
within $0.4''$ of the ACS positions of the 172 
objects in Tables 1 and 2, we calculated median
offsets in RA ($+0.017''$) and dec ($+0.043''$) between COUP
and ACS positions.  We then applied these median offsets to
the ACS positions, so as to refine the search for COUP
counterparts and to calculate the ACS-COUP offset for each
COUP source (this offset is listed under $\Delta_X$ in
Tables 1 and 2). We also correlated COUP-corrected ACS
positions against the near-infrared source positions in the
merged VLT catalog,
resulting in the identifications of near-infrared counterparts
to proplyds listed in Table 1 (where the ACS-infrared positional offset is
given by $\Delta_I$ in Table 1). 

Table 3 summarizes the optical and X-ray properties of {\it
bona fide} (Table 1)  
proplyds having COUP counterparts. The listed stellar spectral
types and visual extinction data were
compiled by Getman et al.\ (2005a), while the X-ray
hardness ratios, absorbing columns ($\log{N_H}$), and
luminosities ($\log{L_{t,c}}$) were
determined by Getman et al.\ from analysis of 
the COUP data. The values for $\log{N_H}$ and
$\log{L_{t,c}}$ have typical formal uncertainties
of $\sim0.1$ dex; however, the systematic
uncertainties --- due to, e.g., assumptions adopted in the
automated spectral modeling procedure --- can be much larger
(see discussion in Getman et al.\ 2005a).

We find the detection
fractions of point-like X-ray and near-infrared counterparts to 
these proplyds are $\sim66$\% and
$\sim92$\%, respectively\footnote{All of the proplyds
listed in Table 1 as not detected (``NV'') in the merged VLT
catalog do have near-infrared counterparts, but these counterparts appear
nebulous rather than point-like.} (Table 4). 
Table 4 makes clear that proplyds with ACS-detected central
stars are detected far more readily in X-rays ($\sim80$\%
COUP detection rate) than proplyds without visible central
stars ($\sim25$\% COUP detection rate). This result, of
course, is most likely due to the larger absorbing
column characteristic of proplyds without optically detected
stars. In \S 4, we elaborate on this result in the context of
proplyds that appear to harbor ``silhouette disks''
(McCaughrean \& O'Dell 1996). 

The high X-ray detection rate of proplyds in the COUP data 
is not surprising, in that it is consistent with the status
of these objects as low-mass, pre-main 
sequence stars embedded in circumstellar disks and/or
envelopes. The detection by COUP of $\sim25$\% of those proplyds that
lack optically detected central stars is 
significant, however, as it indicates that even apparently
``starless'' proplyds in fact harbor optically obscured
PMS stars. 

Though it is not a proplyd, one object in Table 2, 155-040,
is particularly noteworthy. This object is \object{HH 210}, a shocked emission
complex that is found at the tip of one of the ``fingers''
extending radially away from the Kleinmann-Low nebula (Allen
\& Burton 1993). 
COUP X-ray source \object[COUP 703]{703} is found at the apex of a
bow-shock-like structure in HH 210; as such, it is one of
only a handful of
COUP sources thus far identified as counterparts to HH
objects (Getman et al.\ 2005b). The implications of the
detection of X-ray-emitting gas associated with HH 210 will be
considered in detail in Grosso et al.\ (2005, in prep.).

\subsection{Near-infrared colors of proplyds and jet sources}

\begin{figure}
\includegraphics[scale=.9,angle=90]{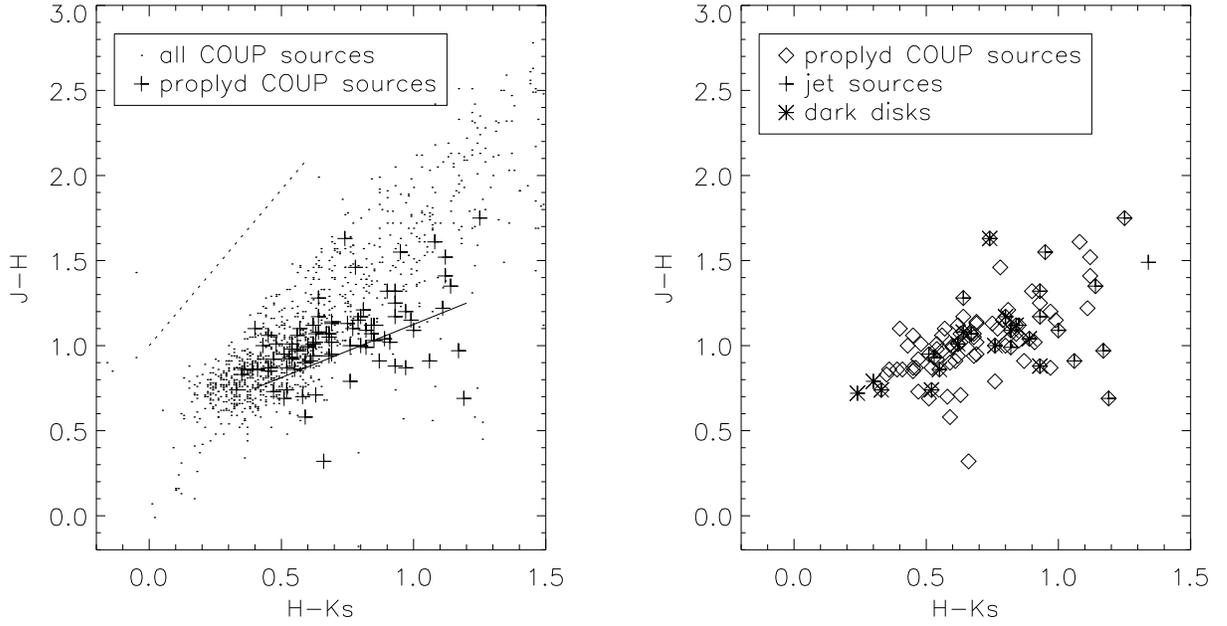} 
\caption{$J-H$ vs.\ $H-K_s$ color-color diagrams for Table 1
objects with available (VLT or 2MASS)
near-infrared photometry. Left: near-infrared colors of Table 1
COUP sources
(crosses) overlaid on a plot of the near-infrared colors of all COUP
sources for which near-infrared photometry is available. The dotted line
indicates the reddening vector for $A_V = 10$, and the solid
line indicates the locus of near-infrared colors of classical T Tauri stars in
Taurus (Meyer et al.\ 1997). Some very heavily absorbed COUP
sources lie off the right edge of the plot. Right: near-infrared colors for 
proplyd sources only, with jet sources (crosses) and silhouette
disk proplyds (asterisks) highlighted (see \S\S 4, 5).} 
\label{fig:nearIR}
\end{figure}

In Fig.~\ref{fig:nearIR} we display a near-infrared
(JHK$_s$) color-color diagram for those COUP sources in
Table 1 for which near-infrared photometry is available via
the VLT imaging or (in a handful of cases) from 2MASS data
(Getman et al.\ 2005a). The plot includes the JHK$_s$ colors
of all COUP sources, for reference. It is apparent that the
proplyd sources, as a class, display red $H-K_s$ colors
relative to $J-H$. Specifically, whereas the vast majority
of the very red COUP X-ray sources lie along the region of
$J-H$ vs.\ $H-K_s$ space that can be attributed to reddening
by intervening dust, the colors of the COUP-detected
proplyds are indicative of near-infrared
excesses. For low-mass,
PMS stars, such excesses are commonly attributed to dusty
accretion disks and, furthermore, the magnitude of infrared
excess appears to be correlated with accretion rate (Meyer
et al.\ 1997). It is therefore noteworthy that the near-infrared
colors of the proplyds resemble those of actively accreting
(classical) T Tauri stars (Fig.~\ref{fig:nearIR}). We
caution that it is possible that Br$\gamma$ emission
from ionized gas at the surfaces, photo-ablation winds,
and/or jets of some proplyds contaminates the $K_s$ band
photometry, mimicking an excess in $H-K_s$. Such an
effect probably is not significant for most sources in
Fig.~\ref{fig:nearIR}, however, as we expect that the total
contribution from diffuse emission typically should be $<1$\% of the
total $K_s$ band flux.

\section{COUP Sources within Silhouette Disk Proplyds}

Certain proplyds exhibit optical absorption morphologies that serve as direct
evidence of the presence of dusty, circumstellar disks (see
discussions in McCaughrean \& O'Dell 1996 and Bally et al.\ 2000). We find 
that a total of 39 proplyds appear to harbor (or consist of)
such silhouette disks (Table 5). Half of these objects have
central stars detected in the visible and/or X-ray 
regimes, and all but 2 silhouette disk proplyds were
detected as point sources in near-infrared 
imaging (both of these ``nondetections,'' d154-240 and d182-413, appear as
extended sources in the VLT images). The latter,
high detection rate confirms that these 
structures are, in fact, circumstellar disks associated
with PMS stars.

ACS images of representative silhouette disks with COUP counterparts (see
below) are displayed in 
Fig.~\ref{fig:darkdisk_imgs}. A histogram of the number of such disks
vs. apparent aspect ratio (i.e., the 
apparent disk major to minor axis ratio, as estimated from
the ACS images; see Table 5) is presented in
Fig.~\ref{fig:histaspect}. The figure indicates that there
the frequency of silhouette disks declines with increasing
aspect ratio. Such a relationship would be expected if these 
objects in fact reflect a population of circularly
symmetric disks viewed at random inclinations. This result confirms
that the aspect ratio of structures seen in silhouette serves as a direct
indicator of disk inclination, as assumed by Bally et al.\
(2000). 

The X-ray detection rate of silhouette disk proplyds is
evidently a steep function of disk 
aspect ratio and, hence, disk inclination (Fig.~\ref{fig:histaspect}).
The effect of disk inclination is also apparent in
representative X-ray spectra of sources associated with
silhouette disks: X-ray sources that are embedded within more
highly inclined silhouette disks generally display harder
spectra that are indicative of larger absorbing columns
(Fig.~\ref{fig:darkdisk_spec}). This effect is shown more clearly
in a plot of $\log{N_{\rm H}}$ vs.\ aspect ratio 
(Fig.~\ref{fig:NHaspect}). This plot demonstrates that
absorbing column increases with increasing aspect ratio, as
would be expected if these structures are in fact dusty
disks, and those objects with aspect ratios $\ge 3$ are
viewed nearly edge-on. Indeed, two out of three silhouette disks that
include embedded X-ray sources and display aspect
ratios $\ge 2.5$ have optically undetected central stars
(Fig.~\ref{fig:NHaspect}). Furthermore, comparison of
Fig.~\ref{fig:histaspect} and Fig.~\ref{fig:NHaspect}
indicates that there is
a systematic bias against detection of X-ray sources in edge-on
(or nearly edge-on) disks, wherein the detection threshold
is $\log{N_{\rm H}}$ (cm$^{-2}$) $\stackrel{>}{\sim} 24$.

A few COUP sources associated with silhouette disks warrant
special attention, as we now describe.  

\subsection{Silhouette disks harboring luminous central stars}

The proplyds d053-717 and d218-354, both of which are
associated with COUP X-ray sources (Table 5), stand out
among the high-aspect-ratio silhouette disks as having
unusually bright central stars in ACS imaging
(Fig.~\ref{fig:darkdisk_imgs}). Indeed, the central star of
d218-354 is saturated in the ACS image. Nevertheless, it
appears that the silhouette disks in both systems are viewed
at large inclination, with d053-717 possibly viewed nearly
edge-on. The absorbing column of $\log N_{\rm H} (\mathrm{cm}^{-2})
= 21.67$ that is associated with \object{COUP 1174}, the X-ray
counterpart to d218-354, is consistent with the value of
$A_V=1.51$ inferred for the coincident optical star, which
is estimated to be of late G or early K type. 

The X-ray counterpart to d053-717, \object{COUP 241},
however, displays a highly
absorbed spectrum (Fig.~\ref{fig:darkdisk_spec})
characterized by $\log N_{\rm H} (\mathrm{cm}^{-2}) = 22.7$. This is
far in excess of the $N_H$ that would be predicted from
the visual absorption of $A_V=0.43$ inferred from
optical/near-infrared photometry of the (mid-K type) central
star, assuming standard ISM values of gas-to-dust ratio. It is
therefore possible that the circumstellar material in 
d053-717 is unusually dust-poor. Alternatively, the
discrepancy could also be explained if, despite 
the apparent positional coincidence, the optically luminous
central star of d053-717 is not the source of X-ray emission
in this system. That is, the X-ray-emitting PMS star may
lie embedded within the disk, with the optically detected
star as binary companion; this would require that the binary
orbit and disk are not coplanar. Finally, it is also possible
that the value of $A_V$ toward the central star of
d053-717 has been underestimated.

\subsection{COUP sources in ``starless'' silhouette disks}

Three COUP sources (\object[COUP 419]{419}, \object[COUP
  476]{476}, and \object[COUP 814]{814}) are 
associated with silhouette disks within which no central stars are
detected in ACS images. We discuss COUP 476 in \S 5. 

COUP source 419 is particularly noteworthy. This weak
source (24 net counts) is the X-ray counterpart of d114-426,
which is among the best examples of an apparent edge-on
silhouette disk (McCaughrean et al.\ 1998;
Fig.~\ref{fig:darkdisk_imgs}). It therefore 
does not appear to be coincidental that the value of
absorbing column determined from the COUP spectrum of this source,
$\log{N_{\rm H}} (\mathrm{cm}^{-2}) = 23.7$, is the largest of any
of the proplyd X-ray counterparts. This source is considered
further in \S 6.1.

The source COUP 814 lies very near the silhouette disk
proplyd 166-519, but may be associated with
a faint, ACS-detected binary companion rather than with the
dark disk.  This interpretation
is supported by the rather modest column density found for
COUP 814 ($\log{N_{\rm H}} (\mathrm{cm}^{-2}) =20.8$) via spectral fitting.

\section{COUP X-ray Detections of Jet Sources in the ONC}

Among the Table 1 sources are 30 objects
exhibiting jet-like structures in either the recently
obtained ACS H$\alpha$
images or in earlier HST narrow-band imagery (Bally et al.\
2000). These objects are listed in Table 6. There is
considerable overlap with the silhouette disk sample; 8 of
these jet sources appear as silhouette 
disks in the ACS images (Tables 5, 6). Approximately
60\% of the jet and ``microjet'' sources were detected in
the COUP X-ray observations (Table 4). Here, we consider in some detail
the X-ray emission properties of one of the more remarkable
examples of a proplyd disk-jet system, d181-825. 

\subsection{The Beehive Proplyd}

The X-ray source \object{COUP 948} is associated with the Beehive
Proplyd, d181-825, which has been described in detail by
Bally et al.\ (2005). This object constitutes one of the most
striking examples of a proplyd disk--jet--ionization front system
(Fig.~\ref{fig:Beehive}). An elliptical silhouette disk is evident
at the center of the object, and jets are observed to protrude
along the minor axis of the ellipse. It is not clear whether
the central star is detected directly in the ACS image,
given the bright jet emission in close proximity to the
apparent position of the central source (perhaps combined
with the enhancement 
of H$\alpha$ emission, relative to the stellar continuum, by the
narrow-band filter used in the 
ACS imaging). Surrounding this
central disk/jet region is an elegant system of ionization fronts
that appears to exhibit a corrugated paraboloid structure. Bally et
al.\ (2005) propose that this structure may
trace density waves moving at about the sound speed
($\sim3$ km s$^{-1}$) in the
neutral medium just inside the ionization fronts.  These
waves likely would be generated by the passage of pulses
of supersonic jet ejecta. Such pulses, in turn, are
responsible for a series of HH objects and bow shocks that
extends several arcmin north and south of d181-825
(Bally et al.\ 2001, 2005). 

The X-ray spectrum of the Beehive source (COUP 948)
stands out among the proplyd X-ray sources.
This spectrum is clearly
double-peaked, consisting of distinct hard and soft
components that correspond to thermal plasma at $kT_1 =
0.57$ keV and $kT_2 = 3.55$ keV
(Fig.~\ref{fig:Beehive}). Furthermore, the results of
spectral fitting by Getman et al.\ (2005a) indicate these two
components are viewed through very different absorbing
columns of $\log{N_{\rm H}} (\mathrm{cm}^{-2})=20.9$ and $\log{N_{\rm H}}
(\mathrm{cm}^{-2})=22.8$, 
respectively. The large absorbing column characterizing the
hard component suggests this emission is strongly attenuated
by gas and dust within the inner regions of the circumstellar
disk detected in the ACS imagery. 
In contrast, the rather small absorbing column toward
the soft component suggests that the soft X-rays are 
subject to little or no attenuation by the circumstellar disk.

To investigate whether the soft X-ray emission is in fact
extended (e.g., is generated in the same shocks that are responsible
for the HH objects in the system), we
carried out spatial deconvolutions of 
the emission from COUP 948. We find that the emission is
point-like, to within the uncertainties. There is
marginal evidence for a small ($<0.2''$) displacement
between the hard and soft components, with the centroid of
the soft component located slightly south of the centroid of the
hard component, in the deconvolved X-ray images. 

The light curves of soft (0.5--2.0 keV) and hard (2.0--8.0
keV) X-ray emission from the Beehive are displayed in
Fig.~\ref{fig:Beehive_lc}. Whereas the light curve of the soft
component is consistent with a constant count rate, the hard
component is clearly variable and displays a strong flare near the
end of the COUP observations.

\subsection{Potential analogs to the Beehive?}

A plot of Chandra/ACIS
hardness ratios as measured in the COUP data is presented in
Fig.~\ref{fig:jetdisk_HRs}, where we include only sources
for which hardness ratio uncertainties are $\le 0.1$ (generally,
this condition is only met by sources with several hundred
counts). This Figure demonstrates that almost all of the 
proplyd X-ray sources lie along a locus of hardness ratios
characteristic of absorbed plasma emission that is dominated
by components with $kT > 1$ keV. However --- as a consequence
of its unusual, double-peaked X-ray spectrum --- COUP 948 lies well above
this locus. 

Only one other well-detected ($>100$ net counts) COUP
proplyd source lies near COUP 948 in 
Fig.~\ref{fig:jetdisk_HRs}. This source, \object{COUP 1011}, is associated
with the proplyd 191-350. Like the Beehive, this
``cometary globule'' proplyd displays well-collimated,
bipolar jets in ACS images
(Fig.~\ref{fig:other_jets}). Although not as clearly 
double-peaked, the X-ray spectrum of COUP 1011 (not shown) 
somewhat resembles that of COUP 948, resulting in its anomalous
hardness ratios.

Among the other COUP counterparts to proplyds that clearly
exhibit jets in ACS imaging (Fig.~\ref{fig:other_jets}), COUP
\object[COUP 476]{476} and \object[COUP 524]{524} (X-ray
counterparts to the jet sources d124-132 and 131-247, 
respectively) also appear to have ``Beehive-like'' hardness
ratios (Table 3). However, while COUP 476 appears
to display a 
double-peaked X-ray spectrum, neither source is well
detected (less than 50 net counts in
each case). This renders their spectral similarity to COUP
948 questionable and, indeed, we have not included these sources in
Fig.~\ref{fig:jetdisk_HRs}. Another five COUP sources
(COUP \object[COUP 279]{279},
\object[COUP 693]{693}, \object[COUP 747]{747}, 
\object[COUP 900]{900}, and \object[COUP 1262]{1262},
        associated with proplyds 069-601, 
152-738, d158-327, 176-325, and 236-527, respectively) also
display anomalous hardness ratios (Table 3). In each case,
however, these sources either suffer from poor photon
counting statistics or high background count rates,
rendering their hardness ratios unreliable (and these
sources therefore are also omitted from
Fig.~\ref{fig:jetdisk_HRs}). Of these five
sources, only COUP 279 --- whose spectrum is dominated by a
soft component characterized by $kT = 0.71$ keV --- appears to
be a viable candidate for shock-generated X-ray
emission. The associated object detected in ACS imaging,
069-600, appears to be surrounded by a wind 
collision front (Bally et al.\ 2000) and 
may be a microjet source 
(Fig.~\ref{fig:other_jets}). We 
discuss these sources in more detail in a forthcoming paper
(Grosso et al.\ 2005, in prep.).

\section{Discussion}

The detection of X-ray emission from a very large
fraction of Orion Nebula Cluster circumstellar disk sources imaged with the
Hubble Space Telescope has a wide range of astrophysical
applications. We focus our discussion here on two
issues. First, the measurement of increasing soft X-ray 
absorption as stellar X-rays penetrate longer path
lengths through circumstellar disks (Figure 5) constrains the geometries and
compositions of such disks (\S 6.1).  Second,  X-ray emission
from jets which power large-scale outflows (\S 5) offers
insight into conditions around the star-disk-outflow
interface (\S 6.2).
 
\subsection{Disk geometry and composition}

The X-ray detections of highly inclined silhouette disks ---
and, in particular, the measurement of the absorbing column
toward COUP 419, the X-ray source detected within the
nearly edge-on disk d114-426 --- are notable in
that they provide unusually clear examples of the X-ray
irradiation of T Tauri accretion disks by the central T
Tauri stars themselves. These results, coupled with the detection in
several COUP sources of fluorescent 6.4 keV line emission
that is evidently due to reflection off of circumstellar
disks (Tsujimoto et al.\ 2005), have a variety of
implications for physical processes in PMS circumstellar disks.
The X-rays absorbed by the disk will affect its ionization,
dynamics (particularly degree of turbulence), heating, and
chemistry, while flare-produced energetic particles may
produce spallogenic nuclear reactions in disk material.
Consideration of these issues lies beyond the scope of this
paper; readers are referred to reviews by Glassgold et al.\
(2000, 2005) and Feigelson (2005).

The X-ray-inferred absorbing columns toward COUP 419 and the
other silhouette proplyds provide, in principle, the first
direct measurements of the gas content of OB photoevaporated
protoplanetary disks.  To explore this potential, which is
manifested in the apparent
relationship between silhouette disk orientation (as
inferred from ACS imaging) and column density (as inferred
from X-ray spectral fitting of COUP counterparts to
silhouette disk proplyds), we employ a simple disk density
model (Aikawa \& Herbst 1999). The Aikawa \& Herbst model
describes the density of H atoms as a function of radial and
vertical displacements within a minimum mass
solar nebula (disk) surrounding a star of solar mass and
luminosity. The spatial scale of this model, which extends
to a radius $R_{\rm out} \sim 10^3$ AU, is compatible with
that of the proplyds in the ACS images. The Aikawa \& Herbst
model is very similar to that formulated by D'Alessio et
al.\ (1999; see also Glassgold et al.\ 2004, their Fig.\
1). The imposition of hydrostatic equilibrium in these
models leads to a flared disk. Such hydrostatic models well
describe accretion disks around T Tauri stars for which the
dominant source of incident radiation is the central star
itself, and for which the gas-to-dust ratios are similar to
those typical of the interstellar medium. In employing the
Aikawa \& Herbst model, we therefore ignore heating and
photoionization due to OB star radiation fields (see, e.g.,
Hollenbach et al. 2000).

We integrated the radial and vertical density distribution of this disk model
over a range of disk inclinations $i$, with inner disk radii
(``holes'') ranging from 0.03 AU (i.e., disk extending
nearly to the PMS stellar photosphere) to 10 AU, to yield
the model dependence of the column density $N_{\rm H}$ on $i$
(Fig.~\ref{fig:modelNHvsi}). To 
facilitate comparison with the observed distribution of 
$N_{\rm H}$ with disk inclination (Fig.~\ref{fig:NHaspect}), we
adopt the disk inclination estimates listed in 
Bally et al.\ (2000) so as to indicate the approximate
positions of three representative proplyds in Fig.~\ref{fig:modelNHvsi}.

This comparison indicates that, for the specific case of d114-426,
the ``canonical'' T Tauri disk model overestimates (by about
two orders of magnitude) the actual proplyd column densities
along (within $\sim10^\circ$ of) the disk plane. However, the same
model also vastly underestimates the column densities for disks d218-354 and
d172-028, which are viewed at
intermediate inclinations. This latter discrepancy
suggests that the scale heights of these ONC proplyds are much
larger than that assumed in the model. This result would
appear to be consistent with the observation that a large
fraction of proplyds (i.e., those noted in Table 1 as ``cometary'' in
appearance) are subject to the intense radiation fields of the Trapezium
OB stars (O'Dell \& Wong 1996; Bally et
al.\ 2000). These fields are rapidly ablating many proplyds, 
resulting in disk mass loss rates that can exceed $10^{-7}$
$M_\odot$ yr$^{-1}$ (St\"orzer \& Hollenbach 1999; Bally et
al.\ 2000). This ablation process is 
responsible for the cometary globule morphologies that are
commonly associated with proplyds, including many of those
contained in the silhouette disk sample (Table 1). One would
therefore expect the scale heights of silhouette disk
proplyds to be systematically larger than those of the ``canonical'' T
Tauri disk model. Such a conclusion is supported by
Fig~\ref{fig:modelNHvsi} (although we note that neither
d218-354 nor d172-028 display clear evidence for ongoing
photo-ablation, in the ACS images). In addition, many silhouette disk
proplyds in Fig.~\ref{fig:NHaspect} likely are viewed
through large intervening absorbing columns (typically
$\sim2\times10^{21}$ cm$^{-2}$, corresponding to $A_V \sim
1$; e.g., O'Dell 2001) due to foreground material,
such that the discrepancies between model and observations
may not be due entirely to the effects of disk
ablation. There is also considerable scatter in the X-ray
absorption measurements shown in Fig.~\ref{fig:NHaspect},
and it is not clear whether this scatter arises from
measurement errors in $\log{N_{\rm H}}$ and disk aspect ratios,
large variations in foreground extinction (O'Dell 2001), or
real differences in disk properties.

The apparent deficit of absorbing material toward the X-ray
source COUP 419 in (apparently edge-on) d114-426 is unlikely
to be caused by photo-ablation of disk gas,
however. Evidently little or no ionizing radiation from the
Trapezium reaches d114-426, as it is not surrounded by an
ionization front. As a consequence, its scale height may be smaller
than those typical of UV-irradiated disks in the ONC,
resulting in a large silhouette aspect ratio at
somewhat more moderate inclination. Given the highly
symmetric optical and near-infrared reflection nebula morphology of 114-426
(McCaughrean et al.\ 1998), however, it seems unlikely that this
proplyd is viewed at an inclination as large as
$\sim10-15^\circ$ with respect to the disk plane (as would be
suggested by the comparison of its measured
$N_{\rm H}$ with the predictions of the simple disk
model; Fig~\ref{fig:modelNHvsi}). 
If the d114-426 silhouette disk is indeed viewed at an
inclination $<10^\circ$,
then its small inferred hydrogen column
density (relative to the model) could be due instead to the depletion 
of neutral metals in the gas phase within the circumstellar
disk. Such an interpretation would be consistent with the
possibility that this silhouette disk harbors a highly evolved
population of large grains (Throop et al.\ 2001; Shuping et
al.\ 2003). Although it is likely that many or even most
circumstellar disks in the ONC are undergoing a similar
process of gas depletion, d114-426 may be somewhat unusual in 
its apparent advanced degree of disk evolution, given the low 
X-ray detection rate of stars within silhouette disks that
are viewed at similar inclinations (Fig.~\ref{fig:histaspect}). 

Intriguingly, the inferred absorbing column toward the
apparent shock zone in the Beehive (COUP 948; \S 5.1) also
points to the possibility of substantial metal depletion in
some proplyd {\it outflows}. The reasoning underlying this
conclusion is as follows.  The presence of a large radius
ionization front surrounding the Beehive Proplyd (and other
large proplyds with H$\alpha$-bright ionization fronts)
indicates that the object is embedded in a dense cocoon. The
contribution of this cocoon to the foreground $N_{\rm H}$ to the
star (and, therefore, the jet collimation and shock zone) is
obtained from the estimated electron density at the
ionization front. The photo-ablation
induced mass-loss rate through such an ionization front
should be $\sim2\times10^{-7}$ $M_\odot$ yr$^{-1}$, assuming
quasi-steady, spherically symmetric outflow at a velocity of
$\sim10$ km s$^{-1}$. This implies an
absorbing column to the disk (i.e., the source of the flow)
of $N_{\rm H} \sim 1.5\times10^{21}$ cm$^{-2}$ for a disk radius
of 50 AU. Given the likelihood that the outflow
speed is even lower, and that we 
view COUP 948 through an additional foreground column, it
thus appears that the modest value of 
absorption toward the soft X-ray-emitting plasma in COUP
948, $N_{\rm H} \approx 8\times10^{20}$ cm$^{-2}$, is best
explained as reflecting the depletion of metals in the
ablating gas.

\subsection{X-ray emission from star-disk-jet interaction regions}

The infrared excesses of many proplyds are detected
shortward of $\sim3$ $\mu$m (Fig.~\ref{fig:nearIR}),
indicating that there exists hot dust quite close to the central
stars in these objects (see also, e.g.,
McCaughrean \& O'Dell 1996; Hayward \& McCaughrean
1997). Indeed, Fig.~\ref{fig:nearIR} provides 
strong evidence that the large-scale, disk-like structures
detected in ACS imaging of proplyds (\S 4) constitute the outer regions of 
circumstellar disks which, in many if not
most cases, extend to within a few stellar radii of the central, PMS
stars. This, in turn, suggests that, for many proplyds, accretion onto the
central star is ongoing. 

Furthermore, the near-infrared
excesses of the jet sources are among the most
extreme exhibited by the COUP-detected proplyd sample
(Fig.~\ref{fig:nearIR}). The magnitude of these excesses suggests 
that the Beehive and other jet sources are actively
accreting at rates of up to $10^{-6}$ $M_\odot$ yr$^{-1}$
(Meyer et al.\ 1997). The inner regions of the accretion
disks in sources such as the Beehive likely provide both 
launching and collimating mechanisms for the observed
jets (Goodson et al.\ 1997; Delamarter et al.\
2000; Matt et al.\ 2003). 

Given this context --- and the scarcity of examples of X-ray
emission from shock-heated gas in collimated protostellar
outflows --- it is therefore quite significant that
at least two COUP counterparts to ONC sources with resolved
optical jets (COUP 948 and 1011)
display unresolved, soft X-ray spectral components
that are indicative of shocks very close to the central stars. 
In particular, the results presented in \S 5.1 concerning COUP 948, the X-ray
counterpart to the central source in the Beehive Proplyd (d181-825),
strongly suggest that its soft X-ray spectral component emanates from
energetic shocks at the base 
of its forward-facing (southeastern-directed) jet. Such an
interpretation is consistent with (1) the relatively low
temperature characteristic of the soft component, (2) the
relatively modest value of $\log{N_{\rm H}} (\mathrm{cm}^{-2}) =20.9$ resulting from
spectral fitting of the two-component plasma model, and (3)
the constant count rate of the soft component. By analogy
with the Beehive, it is likely that the unresolved soft components
in COUP 1011 and, possibly, COUP 279, 476, and 524 also are
generated via energetic shocks in the jet collimation
regions of these PMS star-disk-jet systems.

These definitive and tentative soft X-ray
detections join only a handful of previous X-ray detections
of shock-heated gas in HH outflows (e.g., \object{HH 2},
Pravdo et al.\ 2001; \object{HH 154}, 
Favata et al. 2002, Bally et al.\ 2003; \object{HH 80/81}, Pravdo et
al.\ 2004).  The detection of X-ray-emitting
shocks at the {\it base} of the HH 540 flow from d181-825
appears to be most similar to the case of HH 154 in Taurus (Favata
et al. 2002; Bally et al.\ 2003), and is distinct from HH 2
and HH 80/81, both of which display diffuse X-ray emission
at or near the positions of optical (HH) nebulosity lying far from
the jet sources. Thus,
like the HH 154 X-ray source, the origin of the soft X-rays
from d181-825 may be intimately related to the
jet launching and/or collimation process (see discussion in
Bally et al.\ 2003).

In contrast, the high temperature and variable
flux of the hard X-ray spectral component of COUP 948 indicates that
this component arises in plasma
that is generated via magnetic reconnection events. The large
absorbing column further suggests such events arise deep
within the disk, close to the star. Given that
the X-ray emission from the vast majority of ONC sources
appears to be generated by solar-like coronal activity
(Preibisch et al.\ 2005), a similar mechanism is likely to
be responsible for the hard X-rays observed from COUP
948. However, it is also possible that this hard, variable
X-ray emission may arise from star-disk interactions that
are ultimately responsible for the mass ejections detected
in HST imagery and the shocks detected in soft X-rays. Such
a hard X-ray production mechanism, which has been predicted
theoretically (Hayashi, Shibata, \& Matsumoto 1996), was
also proposed to explain the
coincidence of optical/infrared and X-ray outbursts from V1647 Ori
(Kastner et al.\ 2004; Grosso et al.\ 2005). By extension,
the hard X-ray emission from other proplyds --- particularly
those with large near-infrared excesses --- may be due, in
part, to star-disk interactions.  Regardless of its origin,
however, such high-energy emission will have a profound effect on the
ionization and heating of the inner disk as well as the base
of the outflow (e.g., Shang et al. 2002, 2004; Glassgold et
al. 2004).

\section{Summary}

We have used the very deep (838 ks exposure) COUP
observations of the ONC, as well as deep near-infrared
imaging, to identify and investigate the  
X-ray and near-infrared counterparts to 166 optically
detected objects previously identified as protoplanetary disks
(``proplyds'') and/or optical jet sources. Imaging with HST/ACS provides
improved coordinates and detailed morphologies for all
but a small number of these
proplyds. On the basis of the ACS images, we reject 22
objects as lacking obvious protoplanetary disk, jet, or
globule structures, resulting in a sample of 143
objects (1 object, d347-1535, lies off both the COUP and
near-infrared fields of 
view). Our main results for the 
X-ray and near-infrared  
properties of these objects, and the main conclusions
we draw from these results, are as follows.

\begin{itemize}

\item The vast majority ($\sim70$\%) of proplyds are X-ray
  sources, and an even larger fraction of proplyds ($\ge
  90$\%) reveal central stars in near-infrared imaging. Of
  the $\sim40$ proplyds that do not 
  display central stars in high-resolution ACS imaging, only a
  handful lack both X-ray and near-infrared
  counterparts. The X-ray and near-infrared observations
  presented here therefore establish
  beyond doubt the PMS nature of those proplyds lacking central
  stars in narrow-band optical imaging. 

\item In a near-infrared color-color diagram ($J-H$
  vs. $H-K_s$), most X-ray-emitting ONC proplyds appear to lie
  on or near the locus of points defined by classical T
  Tauri stars in Taurus. Assuming little or no contamination
  of the near-infrared photometry by emission from ionized gas, this
  indicates that the central stars of most proplyds are actively
  accreting T Tauri stars. This result further implies
  that the protoplanetary disk structures detected on scales
  of 10's to 100's of AU via HST imaging are, in fact, the outermost
  regions of accretion disks that extend to 
  within a few stellar radii of the central stars.

\item Almost 40 proplyds appear as (or contain) ``silhouette
  disks'' --- i.e., disk-like structures detected in absorption
  against the bright emission-line background of the Orion
  Nebula in the ACS images --- and $\sim$50\% of these silhouette disks
  harbor X-ray sources. These sources provide clear
  examples of the irradiation of T Tauri star disks by X-rays
  emanating from the central T Tauri stars themselves. Such
  X-ray irradiation likely has a profound
  effect on the heating and chemistry of the inner disk and
  outflow regions surrounding T Tauri stars.

\item For X-ray sources within silhouette disks, we find that X-ray
  absorbing column increases with increasing apparent disk
  inclination. Comparison with a simple model of disk
  density structure suggests that some dusty disks
  surrounding T Tauri stars in the ONC have been inflated by
  the heating and/or ablation resulting from the intense UV
  fields of the Trapezium OB stars. On the other hand, the
  absorbing column inferred 
  toward the X-ray source within the (apparently) nearly
  edge-on disk d114-426 --- albeit the largest such column observed
  among the X-ray-emitting proplyds ($\log{N_{\rm H}}
  (\mathrm{cm}^{-2}) = 23.7$) ---
  is 1--2 orders of magnitude smaller
  than that expected for an edge-on T Tauri disk. This
  suggests that the d114-426 disk has undergone substantial
  gas-phase metal depletion. There is also evidence for 
  metal depletion in the photo-ablation outflow from the
  Beehive Proplyd (d181-825).

\item Approximately 2/3 of the $\sim30$ sources that display
  jets in the ACS images have COUP X-ray counterparts. These
  jet sources display the largest near-infrared excesses
  and, hence, accretion rates among the proplyd X-ray
  sources (again, assuming emission from ionized gas does
  not significantly affect the near-infrared
  photometry). This is consistent with models in which the 
  inner, star-disk interaction regions of accretion disks
  provide the launching as well as collimation mechanisms
  for protostellar jets.

\item One of the most spectacular proplyd jet sources, the
  Beehive (which is the driving source of an extensive
  series of HH objects and associated bow shocks), also
  displays perhaps the most 
  remarkable X-ray spectrum among the proplyds. This
  spectrum is sharply double-peaked, with a lightly 
  absorbed, constant soft component and heavily absorbed,
  variable hard component. We identify a handful of
  additional sources whose X-ray spectra may resemble that
  of the Beehive's X-ray counterpart, COUP 948. We interpret
  the soft X-ray component 
  of COUP 948 (and those of its potential analogs) as
  evidence of the presence of shocked gas 
  at the base of the forward-facing jet. The hard X-ray
  emission in COUP 948 and potential analogs is likely due
  to magnetic reconnection events 
  generated via solar-like coronal
  activity; alternatively, the hard component may emanate
  from the same star-disk interaction regions that are
  responsible for disk launching and collimation.

\end{itemize}

\acknowledgements{COUP is supported by Chandra Guest
   Observer grant SAO GO3-4009A (E. D. Feigelson, PI) as
   well as by the Chandra ACIS Team contract
   NAS8-38252. Additional support for the work described in
   this paper was provided by Chandra
   Guest Observer grant GO4-5012X to RIT.} 

{\it Facilities:} \facility{CXO(ACIS-I)}

\begin{deluxetable}{ccccccccl}
\tabletypesize{\scriptsize}
\tablewidth{0pt}
\tablecaption{Orion Nebula Cluster Proplyds}
\tablehead{
\colhead{Proplyd\tablenotemark{a}}&
\colhead{$\alpha$\tablenotemark{b}}&
\colhead{$\delta$\tablenotemark{b}}&
\colhead{VLT\tablenotemark{c} }&
\colhead{$\Delta_I$\tablenotemark{d}}&
\colhead{COUP\tablenotemark{e} }&
\colhead{$\Delta_X$\tablenotemark{f}}&
\colhead{Star?\tablenotemark{g}} &
\colhead{Appearance in ACS image} \\
 & & & &  ($''$) & &  ($''$) & &
}
\startdata
4596-400 & 05:34:59.56 & -05:24:00.3 &\nodata&\nodata & \object[COUP 137]{137} & 0.06 & Y & I-front\tablenotemark{h} \\
005-514  & 05:35:00.47 & -05:25:14.3 &\nodata&\nodata & \object[COUP 147]{147} & 0.14 & Y & cometary rim, I-front  \\
044-527\tablenotemark{i}  & 05:35:04.43 & -05:25:27.4 &  55 & 0.20 & NC  & \nodata  & N & cometary rim \\
d053-717 & 05:35:05.41 & -05:27:17.2 &\nodata&\nodata & \object[COUP 241]{241} & 0.12 & Y?& dark (edge-on?) disk, companion\\
064-705  & 05:35:06.42 & -05:27:04.7 &\nodata&\nodata & \object[COUP 266]{266} & 0.14 & Y &$0.2''$ sep. double star \\
066-652  & 05:35:06.60 & -05:26:52.0 &\nodata&\nodata & \object[COUP 275]{275} & 0.79 & Y &$0.2''$ sep. double star, bright rim \\
069-600  & 05:35:06.91 & -05:26:00.7 & 131 & 0.26 & \object[COUP 279]{279} & 0.22 & Y & jet?, wind collision front\\
d072-135 & 05:35:07.20 & -05:21:34.4 & 137 & 0.11 & NC  & \nodata  & N & dark disk, cometary rim \\
073-227  & 05:35:07.27 & -05:22:26.6 & 141 & 0.18 & \object[COUP 283]{283} & 0.02 & Y & rim?  \\
097-125  & 05:35:09.68 & -05:21:24.9 & 197 & 0.04 & \object[COUP 336]{336} & 0.31 & Y & rim?  \\
102-233  & 05:35:10.14 & -05:22:32.7 & 214 & 0.18 & \object[COUP 358]{358} & 0.13 & Y & cometary rim \\
102-021  & 05:35:10.19 & -05:20:21.1 & 216 & 0.14 & \object[COUP 362]{362} & 0.17 & Y & cometary rim \\
106-156  & 05:35:10.57 & -05:21:56.3 & 243 & 0.09 & \object[COUP 382]{382} & 0.20 & Y & cometary rim \\
106-417  & 05:35:10.54 & -05:24:16.7 & 242 & 0.09 & \object[COUP 385]{385} & 0.03 & Y?& compact nebula, I-front  \\
d109-247 & 05:35:10.90 & -05:22:46.4 & 261 & 0.14 & \object[COUP 403]{403} & 0.06 & Y & cometary rim \\
d109-327 & 05:35:10.94 & -05:23:26.6 & NV  & \nodata  & NC  & \nodata  & N & dark disk?, cometary rim \\
109-449  & 05:35:10.94 & -05:24:48.7 & 262 & 0.12 & \object[COUP 404]{404} & 0.07 & Y?& compact nebula \\
d110-3035& 05:35:10.99 & -05:30:35.2 &\nodata&\nodata & NC  & \nodata  & N & bipolar jet/nebula\\
d114-426 & 05:35:11.31 & -05:24:26.4 & 277 & 0.14 & \object[COUP 419]{419} & 0.17 & N & dark disk\\
117-025  & 05:35:11.72 & -05:20:25.1 & NV  & \nodata  & NC  & \nodata  & N?& amorphous \\
d117-352 & 05:35:11.72 & -05:23:51.8 & 298 & 0.13 & \object[COUP 443]{443} & 0.12 & N & cometary rim\\
119-340  & 05:35:11.90 & -05:23:39.9 & NV  & \nodata  & NC  & \nodata  & N & cometary rim \\
d121-192 & 05:35:12.09 & -05:19:24.8 &\nodata&\nodata & 460 & 0.25 & Y & dark disk \\
121-434  & 05:35:12.12 & -05:24:33.9 & 316 & 0.10 & \object[COUP 465]{465} & 0.06 & N & cometary rim \\
d124-132 & 05:35:12.38 & -05:21:31.5 & 325 & 0.06 & \object[COUP 476]{476} & 0.16 & N & dark disk, jet?, cometary rim  \\
131-046  & 05:35:13.06 & -05:20:45.9 & 355 & 0.14 & NC  & \nodata  & N & dark disk?, cometary rim \\
131-247  & 05:35:13.10 & -05:22:47.3 & 359 & 0.24 & \object[COUP 524]{524} & 0.11 & N & bright jet, cometary rim \\
d132-042 & 05:35:13.24 & -05:20:41.9 & 366 & 0.15 & NC  & \nodata  & N?& dark disk, jet, cometary rim  \\
d132-183 & 05:35:13.23 & -05:18:33.0 & \nodata&\nodata & NC  & \nodata  & N?& dark disk \\
d135-220 & 05:35:13.51 & -05:22:19.6 & 380 & 0.21 & \object[COUP 551]{551} & 0.11 & Y & cometary rim \\
138-207  & 05:35:13.79 & -05:22:07.1 & 396 & 0.03 & \object[COUP 579]{579} & 0.11 & Y & cometary rim \\
139-320  & 05:35:13.92 & -05:23:20.2 & 405 & 0.09 & \object[COUP 593]{593} & 0.06 & N & cometary rim\\
140-1952 & 05:35:14.05 & -05:19:52.1 & 410 & 0.13 & \object[COUP 597]{597} & 0.12 & Y & dark halo \\
d141-520 & 05:35:14.05 & -05:25:20.4 & 411 & 0.12 & \object[COUP 604]{604} & 0.14 & Y & dark disk, cometary rim  \\
d141-301 & 05:35:14.15 & -05:23:01.1 & \nodata & \nodata  & NC  & \nodata  & Y?&(saturated?), cometary rim, dark interior \\
143-425  & 05:35:14.26 & -05:24:24.8 & 422 & 0.15 & \object[COUP 616]{616} & 0.20 & Y & I-front, no proplyd? \\
d143-522 & 05:35:14.34 & -05:25:22.2 & 427 & 0.29 & NC  & \nodata  & N & dark disk, cometary rim \\
144-334  & 05:35:14.38 & -05:23:33.7 & 432 & 0.25 & \object[COUP 631]{631} & 0.20 & Y & I-front?, no proplyd \\
d147-323 & 05:35:14.72 & -05:23:23.0 & 453 & 0.16 & \object[COUP 658]{658} & 0.11 & Y & dark disk, cometary rim \\
150-231  & 05:35:15.02 & -05:22:31.1 & 476 & 0.19 & \object[COUP 678]{678}?& 1:   & Y & cometary rim \\
152-319  & 05:35:15.20 & -05:23:18.9 & 490 & 0.19 & \object[COUP 690]{690} & 0.11 & N & cometary rim \\
152-738  & 05:35:15.21 & -05:27:37.8 &\nodata&\nodata & \object[COUP 693]{693} & 0.45 & Y & I-front? \\
153-1902 & 05:35:15.35 & -05:19:02.2 & \nodata& \nodata  & \object[COUP 695]{695} & 0.13 & Y & compact nebula \\
154-324  & 05:35:15.35 & -05:23:24.2 & 502 & 0.19 & NC  & \nodata  & Y & jet, no proplyd \\
154-225  & 05:35:15.37 & -05:22:25.4 & 505 & 0.21 & \object[COUP 699]{699} & 0.17 & Y & cometary rim \\
d154-240 & 05:35:15.38 & -05:22:40.0 & NV  & ...  & NC  & \nodata  & N & dark disk?, cometary rim  \\
d155-338 & 05:35:15.52 & -05:23:37.5 & 513 & 0.04 & \object[COUP 717]{717} & 0.11 & Y &(saturated), cometary rim \\
156-403  & 05:35:15.61 & -05:24:03.1 & 522 & 0.10 & \object[COUP 726]{726} & 0.08 & Y &(saturated), no proplyd? \\
157-533  & 05:35:15.67 & -05:25:33.1 & 525 & 0.13 & \object[COUP 728]{728} & 0.16 & Y &(saturated), cometary rim \\
157-323  & 05:35:15.72 & -05:23:22.5 & 531 & 0.12 & \object[COUP 733]{733} & 0.12 & Y &(saturated) \\
d158-326 & 05:35:15.79 & -05:23:26.7 & 537 & 0.12 & NC  & \nodata  & Y &(saturated), cometary tail \\
d158-327 & 05:35:15.84 & -05:23:25.6 & 543 & 0.11 & \object[COUP 747]{747} & 0.07 & Y &(saturated), cometary tail \\
158-323  & 05:35:15.83 & -05:23:22.5 & 542 & 0.25 & \object[COUP 746]{746} & 0.13 & Y &(saturated), cometary tail  \\
159-338  & 05:35:15.90 & -05:23:38.0 & 549 & 0.04 & \object[COUP 757]{757} & 0.12 & Y &(saturated), cometary tail  \\
d159-418 & 05:35:15.90 & -05:24:17.8 & 550 & 0.13 & \object[COUP 748]{748} & 0.88 & N & cometary rim\\
159-350  & 05:35:15.95 & -05:23:50.0 & 551 & 0.04 & \object[COUP 758]{758} & 0.15 & Y &(saturated), cometary rim \\
160-353  & 05:35:16.00 & -05:23:53.1 & 559 & 0.15 & \object[COUP 768]{768} & 0.15 & Y &(saturated), cometary rim \\
161-324  & 05:35:16.06 & -05:23:24.4 & 564 & 0.19 & NC  & \nodata  & Y &(saturated), cometary tail \\
d161-328 & 05:35:16.07 & -05:23:27.9 & 566 & 0.17 & NC  & \nodata  & Y?& cometary tail \\
161-314  & 05:35:16.10 & -05:23:14.3 & 571 & 0.17 & \object[COUP 779]{779} & 0.18 & Y?& fuzzy \\
163-317  & 05:35:16.28 & -05:23:16.6 & 585 & 0.20 & \object[COUP 787]{787} & 0.03 & Y &(saturated), cometary tail  \\
d163-026\tablenotemark{j} & 05:35:16.29 & -05:20:25.5 & 588 & 0.27 & NC  & \nodata  & N & dark disk, binary? \\
d163-222 & 05:35:16.30 & -05:22:21.6 & 590 & 0.26 & \object[COUP 799]{799} & 0.27 & Y & dark disk, cometary rim \\
163-249  & 05:35:16.33 & -05:22:49.1 & 592 & 0.25 & \object[COUP 800]{800} & 0.16 & Y & cometary tail \\
164-511  & 05:35:16.36 & -05:25:09.6 & 593 & 0.07 & \object[COUP 803]{803} & 0.16 & Y & jet?\\
165-235  & 05:35:16.48 & -05:22:35.2 & 602 & 0.23 & \object[COUP 807]{807} & 0.12 & Y & cometary rim \\
d165-254 & 05:35:16.54 & -05:22:53.7 & 605 & 0.16 & NC  & \nodata  & N & dark disk \\
166-519  & 05:35:16.58 & -05:25:17.7 & 607 & 0.08 & \object[COUP 814]{814}\tablenotemark{k} & 0.18 & N?& dark disk?, binary? \\
166-250  & 05:35:16.59 & -05:22:50.4 & NV  & \nodata  & NC  & \nodata  & N & cometary tail \\
166-316  & 05:35:16.61 & -05:23:16.2 & 611 & 0.22 & \object[COUP 820]{820} & 0.17 & Y &(saturated) \\
d167-231 & 05:35:16.73 & -05:22:31.3 & 618 & 0.20 & \object[COUP 825]{825} & 0.04 & Y & dark disk  \\
167-317  & 05:35:16.75 & -05:23:16.2 & 619 & 0.18 & \object[COUP 826]{826} & 0.24 & Y &(saturated), cometary tail? \\
168-328  & 05:35:16.76 & -05:23:28.1 & 622 & 0.22 & \object[COUP 827]{827} & 0.06 & Y &(saturated), cometary tail \\
168-235  & 05:35:16.83 & -05:22:34.6 & NV  & \nodata  & NC  & \nodata  & N & cometary rim \\
168-326  & 05:35:16.84 & -05:23:26.3 & 626 & 0.17 & NC  & \nodata  & Y &(saturated), cometary tail \\
169-338  & 05:35:16.88 & -05:23:38.1 & NV  & \nodata  & NC  & \nodata  & Y & cometary tail \\
d170-249 & 05:35:16.97 & -05:22:48.7 & 638 & 0.21 & \object[COUP 844]{844} & 0.22 & N & cometary rim\\
170-337  & 05:35:17.00 & -05:23:37.1 & 640 & 0.17 & \object[COUP 847]{847} & 0.36 & Y &(saturated), cometary tail \\
d171-340 & 05:35:17.05 & -05:23:39.8 & 644 & 0.21 & \object[COUP 856]{856} & 0.10 & Y & cometary rim \\
171-334  & 05:35:17.06 & -05:23:34.1 & 645 & 0.21 & \object[COUP 855]{855} & 0.05 & Y &(saturated) \\
d172-028 & 05:35:17.22 & -05:20:27.7 & 654 & 0.11 & \object[COUP 865]{865} & 0.10 & Y & dark disk? \\
173-341  & 05:35:17.32 & -05:23:41.5 & 658 & 0.22 & NC  & \nodata  & Y & cometary tail \\
d174-236 & 05:35:17.34 & -05:22:35.8 & 661 & 0.32 & \object[COUP 876]{876} & 0.14 & Y &(saturated), cometary rim \\
174-414  & 05:35:17.39 & -05:24:13.7 & 668 & 0.20 & \object[COUP 887]{887} & 0.29 & Y & cometary tail \\
175-251  & 05:35:17.48 & -05:22:51.4 & 674 & 0.18 & \object[COUP 884]{884} & 0.10 & Y & cometary tail  \\
d175-355 & 05:35:17.54 & -05:23:55.1 & 681 & 0.11 & NC  & \nodata  & N & compact rim\\
d176-543 & 05:35:17.55 & -05:25:42.7 & 679 & 0.19 & \object[COUP 901]{901} & 0.13 & Y & dark disk, jet, cometary rim \\
176-325  & 05:35:17.55 & -05:23:24.9 & 683 & 0.28 & \object[COUP 900]{900} & 0.22 & Y &(saturated), cometary rim \\
d177-341 & 05:35:17.67 & -05:23:40.9 & 693 & 0.23 & NC  & \nodata  & Y &(saturated), cometary rim \\
177-454  & 05:35:17.69 & -05:24:53.9 & 694 & 0.16 & \object[COUP 914]{914} & 0.25 & Y & bright rim \\
d177-541 & 05:35:17.71 & -05:25:40.8 & 695 & 0.14 & NC  & \nodata  & N & dark disk, cometary rim \\
177-444  & 05:35:17.73 & -05:24:43.6 & 697 & 0.11 & NC  & \nodata  & Y & cometary tail \\
d179-353 & 05:35:17.96 & -05:23:53.6 & 720 & 0.16 & NC  & \nodata  & N & cometary tail \\
180-331  & 05:35:18.04 & -05:23:30.8 & 726 & 0.19 & NC  & \nodata  & N & cometary tail \\
d181-247 & 05:35:18.08 & -05:22:47.1 & 729 & 0.20 & NC  & \nodata  & N & dark disk, cometary tail \\
d181-825 & 05:35:18.10 & -05:28:25.0 &\nodata&\nodata & \object[COUP 948]{948} & 0.15 & N?& jet, dark disk, I front; Beehive Proplyd\\
d182-332 & 05:35:18.19 & -05:23:31.6 & 731 & 0.16 & NC  & \nodata  & Y?& dark disk  \\
d182-413 & 05:35:18.22 & -05:24:13.4 & NV  & \nodata  & NC  & \nodata  & N & dark disk, cometary rim \\
182-316  & 05:35:18.24 & -05:23:15.7 & 738 & 0.21 & \object[COUP 955]{955} & 0.12 & Y & cometary tail? \\
183-439  & 05:35:18.28 & -05:24:38.7 & 739 & 0.20 & NC  & \nodata  & Y & cometary tail; faint companion\\
d183-419 & 05:35:18.31 & -05:24:18.8 & 743 & 0.20 & NC  & \nodata  & N & dark disk, cometary rim  \\
d183-405 & 05:35:18.33 & -05:24:04.7 & 746 & 0.17 & \object[COUP 966]{966} & 0.22 & Y & dark disk \\
184-427  & 05:35:18.35 & -05:24:26.7 & 749 & 0.27 & \object[COUP 967]{967} & 0.36 & Y & cometary rim; faint companion \\
184-520  & 05:35:18.45 & -05:25:19.2 & 753 & 0.13 & NC  & \nodata  & Y & cometary rim, nebulosity \\
189-329  & 05:35:18.87 & -05:23:28.9 & 776 & 0.16 &\object[COUP 1000]{1000} & 0.01 & Y & cometary tail? \\
191-350  & 05:35:19.07 & -05:23:49.5 & 788 & 0.15 &\object[COUP 1011]{1011} & 0.27 & Y & jet?, nebulosity \\
d191-232 & 05:35:19.125& -05:22:31.4 & 794 & 0.16 & NC  & \nodata  & N & dark disk \\
d197-427 & 05:35:19.66 & -05:24:26.4 & 813 & 0.23 &\object[1045]{1045} & 0.27 & Y & dark disk, cometary rim \\
198-222  & 05:35:19.82 & -05:22:21.6 & 819 & 0.18 &\object[COUP 1056]{1056} & 0.09 & Y & cometary rim \\
198-448  & 05:35:19.84 & -05:24:47.8 & 820 & 0.17 &\object[COUP 1058]{1058} & 0.17 & Y & cometary rim  \\
201-534  & 05:35:20.15 & -05:25:33.7 & 839 & 0.19 & NC  & \nodata  & Y & jet? \\
202-228  & 05:35:20.15 & -05:22:28.3 & 840 & 0.13 &\object[COUP 1084]{1084} & 0.07 & Y & dark disk, cometary rim  \\
d203-504 & 05:35:20.27 & -05:25:03.9 & 847 & 0.24 &\object[COUP 1091]{1091} & 0.10 & Y & cometary rim \\
d203-506 & 05:35:20.32 & -05:25:05.5 & 849 & 0.22 & NC  & \nodata  & N & dark disk  \\
205-330  & 05:35:20.46 & -05:23:29.7 & 855 & 0.38 &\object[COUP 1101]{1101} & 0.18 & Y & cometary rim; companion? \\
205-052  & 05:35:20.52 & -05:20:52.1 & 859 & 0.08 &\object[COUP 1104]{1104} & 0.04 & Y & cometary rim \\
d205-421 & 05:35:20.54 & -05:24:20.8 & 861 & 0.15 &\object[COUP 1107]{1107} & 0.15 & Y & dark disk, cometary rim \\
d206-446 & 05:35:20.63 & -05:24:46.3 & 864 & 0.24 &\object[COUP 1112]{1112} & 0.22 & Y & dark disk, cometary rim \\
208-122  & 05:35:20.84 & -05:21:21.5 & 876 & 0.14 &\object[COUP 1120]{1120} & 0.02 & Y & jet? \\
212-557  & 05:35:21.16 & -05:25:56.9 & 892 & 0.26 &\object[COUP 1139]{1139} & 0.19 & Y & irreg. nebula \\
212-260  & 05:35:21.24 & -05:22:59.5 & 896 & 0.14 &\object[COUP 1141]{1141} & 0.04 & Y & cometary tail \\
213-346  & 05:35:21.31 & -05:23:46.0 & 901 & 0.06 &\object[COUP 1149]{1149} & 0.90 & Y & cometary tail \\
215-317  & 05:35:21.51 & -05:23:16.6 & 907 & 0.14 &\object[COUP 1155]{1155} & 0.27 & \nodata& [off ACS FOV] \\
d216-0939& 05:35:21.57 & -05:09:38.9 &\nodata&\nodata &\nodata&\nodata & ..& [off ACS FOV] \\
218-339  & 05:35:21.77 & -05:23:39.2 & 920 & 0.04 &\object[COUP 1167]{1167} & 0.21 & Y & cometary tail \\
d218-354 & 05:35:21.81 & -05:23:53.7 & 922 & 0.09 &\object[COUP 1174]{1174} & 0.27 & Y & dark disk  \\
d218-529 & 05:35:21.83 & -05:25:28.3 & 924 & 0.25 & NC  & \nodata  & N & dark disk, jet, cometary rim?\\
221-433  & 05:35:22.09 & -05:24:32.7 & 932 & 0.19 &\object[COUP 1184]{1184} & 0.16 & Y & cometary rim \\
224-728  & 05:35:22.38 & -05:27:28.3 &\nodata&\nodata&\object[COUP 1206]{1206} & 0.13 & Y & cometary rim \\
228-548  & 05:35:22.83 & -05:25:47.5 & 968 & 0.26 & NC  & \nodata  & Y & cometary tail \\
231-502  & 05:35:23.16 & -05:25:02.2 & 978 & 0.20 & NC  & \nodata  & Y?& compact nebula \\
232-453  & 05:35:23.21 & -05:24:52.8 & 983 & 0.32 & NC  & \nodata  & ..& on CCD bleed artifact\\
236-527  & 05:35:23.60 & -05:25:26.4 &1000 & 0.22 &\object[COUP 1262]{1262} & 0.19 & Y & cometary tail \\
237-627  & 05:35:23.66 & -05:26:27.0 &1004 & 0.22 &\object[COUP 1263]{1263} & 0.25 & Y & cometary rim  \\
d239-334\tablenotemark{j} & 05:35:23.87 & -05:23:34.0 &1015 & 0.78 & NC  & \nodata  & ..& [off ACS FOV] \\
239-510  & 05:35:23.98 & -05:25:09.8 &1021 & 0.20 &\object[COUP 1275]{1275} & 0.18 & Y & compact nebula  \\
240-314  & 05:35:24.05 & -05:23:13.8 &1022 & 0.32 &\object[COUP 1276]{1276} & 0.42 & ..& [off ACS FOV] \\
242-519  & 05:35:24.26 & -05:25:18.6 &1027 & 0.21 &\object[COUP 1281]{1281} & 0.20 & Y & cometary tail  \\
d244-440 & 05:35:24.44 & -05:24:39.8 &1033 & 0.07 &\object[COUP 1290]{1290} & 0.15 & Y & giant cometary proplyd \\
245-632  & 05:35:24.46 & -05:26:31.4 &1035 & 0.23 &\object[COUP 1291]{1291} & 0.04 & Y & cometary rim \\
245-502  & 05:35:24.51 & -05:25:01.5 &1038 & 0.14 &\object[COUP 1293]{1293} & 0.44 & Y & cometary tail \\
247-436  & 05:35:24.70 & -05:24:35.6 &1046 & 0.13 &\object[COUP 1302]{1302} & 0.18 & Y & cometary rim, jet \\
250-439  & 05:35:25.03 & -05:24:38.4 &1055 & 0.13 &\object[COUP 1313]{1313} & 0.12 & Y & cometary tail \\
d252-457 & 05:35:25.21 & -05:24:57.2 &1060 & 0.16 &\object[COUP 1317]{1317} & 0.24 & Y & cometary rim, jet \\
d253-1536\tablenotemark{j}& 05:35:25.30 & -05:15:35.5 &\nodata&\nodata & NC  & \nodata  & Y?& dark disk, jets \\
264-532  & 05:35:26.42 & -05:25:31.5 &1096 & 0.19 & NC  & \nodata  &\nodata & [off ACS FOV] \\
d280-1720& 05:35:28.04 & -05:17:20.2 &\nodata&\nodata &\object[COUP 1404]{1404} & 0.24 & Y & dark disk \\
282-458  & 05:35:28.21 & -05:24:58.2 &1144 & 0.01 &\object[COUP 1409]{1409} & 0.15 & \nodata & [off ACS FOV] \\
d294-606 & 05:35:29.48 & -05:26:06.6 &1164 & 0.11 & NC  & \nodata  & N & dark disk\\
d347-1535& 05:35:34.67 & -05:15:34.8 &\nodata&\nodata &\nodata&\nodata & N & dark disk, bipolar jet\\
\enddata

\tablenotetext{a}{Proplyd candidates with prefix ``d''
are from lists in Bally et al.\ (2000) or Smith et al.\
(2004); all other candidates are from O'Dell \& Wong (1996).}
\tablenotetext{b}{J2000 coordinates determined from ACS images; for sources outside ACS FOV,
  coordinates are from Bally et al.\ (2000), Smith et al.\
 (2004), or O'Dell \& Wong (1996). See text.}
\tablenotetext{c}{VLT IR source number. The text ``NV''
  indicates no IR counterpart detected; ellipsis 
  indicate candidates lying outside the VLT image FOV.}
\tablenotetext{d}{Offset (arcsec) between visual position
(as determined from ACS image) and infrared source position.}
\tablenotetext{e}{COUP X-ray source number.  The text
  ``NC'' indicate no
  X-ray counterpart detected; ellipsis 
  indicate candidates lying outside the COUP FOV.}
\tablenotetext{f}{Offset (arcsec) between visual position
(as determined from ACS image) and X-ray source position.}
\tablenotetext{g}{Y = star apparent in ACS image; N = no star apparent in ACS image}
\tablenotetext{h}{``I-front'': ionization front apparent.}
\tablenotetext{i}{Proplyd 044-527 is a new identification.}
\tablenotetext{j}{d163-026s is found $\sim0.4''$ from
COUP 796 (= MLLA-956) and d239-334 is found $\sim0.6''$ from
COUP 1268 (= MLLA-347), but optical/IR sources near these
proplyds are in fact the X-ray sources; see Fig.\ 7 of Bally
et al.\ (2000). In addition, COUP 1316 lies at a companion
to d253-1536.}  
\tablenotetext{k}{COUP 814 lies very near the position of
  166-519, but may instead be associated with a
  companion. See \S 4.2.}

\end{deluxetable}

{
\begin{deluxetable}{cccccl}
\tablewidth{0pt}
\tablecaption{Objects Not Considered Proplyds}
\tablehead{
\colhead{Object}&
\colhead{$\alpha$\tablenotemark{a}}&
\colhead{$\delta$\tablenotemark{a}}&
\colhead{COUP\tablenotemark{b} }&
\colhead{$\Delta$\tablenotemark{c}}&
\colhead{Comments} \\
 & & & & ($''$) &
}
\startdata
113-153  & 05:35:11.35 & -05:21:53.1 & ... & ...  &HH knot(s) \\
114-155  & 05:35:11.44 & -05:21:54.8 & ... & ...  &HH knot(s) \\
115-155  & 05:35:11.55 & -05:21:54.5 & ... & ...  &HH knot(s) \\
116-156  & 05:35:11.59 & -05:21:55.8 & ... & ...  &HH knot(s) \\
127-711  & 05:35:12.71 & -05:27:10.7 & \object[COUP 498]{498} & 0.05 &double star \\
128-044  & 05:35:12.81 & -05:20:43.6 & \object[COUP 501]{501} & 0.06 &triple star  \\
132-221  & 05:35:13.17 & -05:22:21.3 & \object[COUP 523]{523} & 0.20 &double star \\
132-222  & 05:35:13.26 & -05:22:21.8 & ... & ...  &no source in ACS image \\
135-227  & 05:35:13.47 & -05:22:27.0 & ... & ...  &no source in ACS image \\
137-222  & 05:35:13.73 & -05:22:22.0 & \object[COUP 573]{573} & 0.14 &star \\
144-522  & 05:35:14.41 & -05:25:21.5 & ... & ...  &no source in ACS image\\
149-329  & 05:35:14.92 & -05:23:29.1 & \object[COUP 671]{671} & 0.20 &star (saturated?) \\
153-321  & 05:35:15.35 & -05:23:21.4 & ... & ...  &star  \\
154-042  & 05:35:15.45 & -05:20:41.9 & ... & ...  &HH knot(s) \\
155-040  & 05:35:15.49 & -05:20:40.1 & \object[COUP 703]{703} & n/a  &HH knot(s) \\
158-425  & 05:35:15.77 & -05:24:24.8 & \object[COUP 736]{736} & 0.08 &star \\
169-549  & 05:35:16.90 & -05:25:49.1 & ... & ...  &no source in ACS image \\
172-327  & 05:35:17.22 & -05:23:26.7 & ... & ...  &part of WC front? \\
174-400  & 05:35:17.38 & -05:24:00.3 & \object[COUP 880]{880} & 0.10 &star \\
179-536  & 05:35:17.90 & -05:25:35.9 & ... & ...  &no source in ACS image \\
187-314  & 05:35:18.66 & -05:23:14.0 & \object[COUP 986]{986} & 0.05 &double star \\
222-637  & 05:35:22.20 & -05:26:37.4 & \object[COUP 1202]{1202} & 0.11 &double star \\
\enddata

\tablenotetext{a}{J2000 coordinates, as determined from ACS images.}
\tablenotetext{b}{COUP source number.}
\tablenotetext{c}{Offset (arcsec) between visual position
(as determined from ACS image) and COUP X-ray source position.}

\end{deluxetable}
}

{
\begin{deluxetable}{cccccrrrcc}
\tabletypesize{\scriptsize}
\tablewidth{0pt}
\tablecaption{Proplyds with COUP Counterparts: Optical and X-ray Properties} 
\tablehead{
\colhead{Proplyd}&
\colhead{Sp.\ Type\tablenotemark{a}}&
\colhead{$A_V$}&
\colhead{COUP}&
\colhead{Exp.\tablenotemark{b}}&
\colhead{Counts\tablenotemark{c}}&
\colhead{HR2\tablenotemark{d}}&
\colhead{HR3\tablenotemark{e}} &
\colhead{$\log{N_H}$\tablenotemark{f}} &
\colhead{$\log{L_{t,c}}$\tablenotemark{g}} \\
 & & (mag) & & (ks) & &  &  & (cm$^{-2}$) & (erg s$^{-1}$)
}
\startdata
  4596-400   &  M2.5-M4 &  1.03 &  137 & 817.0  &  501 & $-0.15\pm$0.05 & $-0.16\pm$0.06 &   22.0 & 29.6 \\
   005-514   & K6e &  0.48 &  147 & 779.8  & 2348 & $-0.60\pm$0.02 & $-0.34\pm$0.04 &   21.3 & 29.9 \\
  d053-717   &K5-K6 &  0.43 &  241 & 806.4  &  314 & $ 0.65\pm$0.08 & $ 0.30\pm$0.06 &   22.7 & 29.9 \\
   064-705   &\nodata  &   \nodata &  266 & 760.4  &  831 & $-0.75\pm$0.02 & $-0.36\pm$0.08 &   21.2 & 29.4 \\
   066-652   &    M4.5e &  0.47 &  275 & 813.4  &  134 & $-0.76\pm$0.06 & $-0.97\pm$0.33 &   21.7 & 29.0 \\
   069-601   & \nodata  &   \nodata &  279 & 565.9  &   34 & $-0.95\pm$0.11 & $ 0.64\pm$0.68 &   21.9 & 28.6 \\
   073-227   &   M2-M4: &  0.60 &  283 & 778.1  & 3785 & $-0.58\pm$0.01 & $-0.19\pm$0.03 &   21.1 & 30.1 \\
   097-125   &     M3.5 &   \nodata &  336 & 822.3  &  309 & $-0.74\pm$0.07 & $-0.13\pm$0.22 &   21.6 & 29.1 \\
   102-233   &\nodata  &   \nodata &  358 & 344.8  &   93 & $ 0.33\pm$0.12 & $-0.25\pm$0.12 &   22.6 & 30.0 \\
   102-021   &     M3.5 &   \nodata &  362 & 809.9  &  220 & $-0.38\pm$0.07 & $-0.40\pm$0.12 &   21.9 & 29.2 \\
   106-156   &    K2-M2 &  0.42 &  382 & 831.1  & 5065 & $-0.63\pm$0.01 & $-0.38\pm$0.03 &   21.3 & 30.2 \\
   106-417   &       K: &   \nodata &  385 & 838.2  &  885 & $ 0.63\pm$0.04 & $ 0.36\pm$0.03 &   22.6 & 30.2 \\
  d109-247   &  mid-K:: &   \nodata &  403 & 334.2  &  456 & $ 0.47\pm$0.07 & $ 0.36\pm$0.05 &   22.5 & 30.2 \\
   109-449   &      M3e &   \nodata &  404 & 838.2  & 2413 & $-0.44\pm$0.02 & $-0.27\pm$0.03 &   21.7 & 30.1 \\
  d114-426   &\nodata  &   \nodata &  419 & 831.1  &   24 &    \nodata         & $ 0.96\pm$0.15 &   23.7 & 30.2 \\
  d117-352   &\nodata  &   \nodata &  443 & 785.1  &   32 & $ 0.23\pm$0.26 & $-0.04\pm$0.23 &   21.8 & 28.3 \\
  d121-192   &     M4.5 &  5.83 &  460 & 799.3  &  349 & $-0.42\pm$0.05 & $-0.44\pm$0.09 &   21.6 & 29.2 \\
   121-434   &\nodata  &   \nodata &  465 & 772.8  &   35 & $-0.32\pm$0.19 & $-0.22\pm$0.32 &   21.6 & 28.2 \\
   124-132   &\nodata  &   \nodata &  476 & 827.6  &   47 & $-0.68\pm$0.23 & $ 0.71\pm$0.22 &   20.0 & 28.2 \\
   131-247   &       K: &   \nodata &  524 & 834.7  &   32 & $-0.19\pm$0.34 & $ 0.57\pm$0.19 &   22.4 & 28.5 \\
  d135-220   &     M1.4 &  1.34 &  551 & 832.9  &  654 & $-0.56\pm$0.03 & $-0.42\pm$0.07 &   21.6 & 29.5 \\
   138-207   &   K2e-M4 &   \nodata &  579 & 831.1  & 7100 & $-0.32\pm$0.01 & $-0.04\pm$0.02 &   21.7 & 30.6 \\
   139-320   &\nodata  &   \nodata &  593 & 578.3  &  329 & $-0.64\pm$0.05 & $-0.34\pm$0.12 &   21.3 & 29.2 \\
  140-1952   &   late-G &  2.69 &  597 & 804.6  & 8523 & $-0.73\pm$0.01 & $-0.33\pm$0.02 &   20.0 & 30.3 \\
  d141-520   &\nodata  &   \nodata &  604 & 834.7  &  305 & $-0.52\pm$0.05 & $-0.51\pm$0.10 &   21.2 & 28.9 \\
   143-425   &    K4-M1 &  1.24 &  616 & 535.8  & 1096 & $-0.63\pm$0.02 & $-0.38\pm$0.06 &   21.3 & 29.7 \\
   144-334   &       M1 &  1.54 &  631 & 452.7  & 1990 & $-0.52\pm$0.02 & $-0.33\pm$0.04 &   21.5 & 30.1 \\
  d147-323   &      M3e &   \nodata &  658 & 834.7  & 2541 & $ 0.18\pm$0.03 & $ 0.03\pm$0.02 &   22.2 & 30.4 \\
   150-231   &\nodata  &   \nodata &  678 & 832.9  &  614 & $ 0.63\pm$0.15 & $ 0.81\pm$0.03 &   22.9 & 30.3 \\
   152-319   &\nodata  &   \nodata &  690 & 834.7  &   71 & $ 0.52\pm$0.29 & $ 0.56\pm$0.11 &   22.7 & 29.3 \\
   152-738   &\nodata  &   \nodata &  693 & 806.4  &   38 & $-0.43\pm$0.47 & $ 0.85\pm$0.14 &   23.2 & 29.5 \\
  153-1902   &M4.5-M5.5 &   \nodata &  695 & 792.2  &  637 & $-0.76\pm$0.03 & $-0.55\pm$0.12 &   20.0 & 29.1 \\
   154-225   &    M0-M3 &   \nodata &  699 & 831.1  &  370 & $-0.42\pm$0.05 & $-0.41\pm$0.08 &   21.6 & 29.5 \\
  d155-338   &\nodata  &   \nodata &  717 & 836.4  &   93 & $-0.39\pm$0.11 & $-0.37\pm$0.19 &   21.0 & 28.5 \\
   156-403   &    K8-M0 &  0.81 &  726 & 295.3  &  466 & $-0.68\pm$0.04 & $-0.40\pm$0.10 &   21.4 & 29.6 \\
   157-533   &   K8e-M0 &  1.71 &  728 & 832.9  &  190 & $-0.04\pm$0.09 & $-0.04\pm$0.09 &   22.0 & 29.1 \\
   157-323   &\nodata  &   \nodata &  733 & 834.7  &   41 & $ 0.51\pm$0.53 & $-0.13\pm$0.24 &   22.5 & 29.4 \\
  d158-327   &\nodata  &   \nodata &  747 & 834.7  &   48 & $-0.20\pm$0.48 & $ 0.62\pm$0.22 &   22.3 & 28.9 \\
   158-323   & K1-midKe &   \nodata &  746 & 834.7  &   52 & $-0.08\pm$0.35 & $ 0.14\pm$0.26 &   22.1 & 28.8 \\
   159-338   &\nodata  &   \nodata &  757 & 836.4  &   25 & $-0.02\pm$0.29 & $-0.26\pm$0.33 &   22.5 & 29.0 \\
   159-350   &   G5-K0e &  3.78 &  758 & 788.7  &20627 & $-0.31\pm$0.01 & $-0.15\pm$0.01 &   21.7 & 31.1 \\
   160-353   &   F2-F7e &  4.33 &  768 & 687.9  & 1530 & $-0.33\pm$0.03 & $-0.18\pm$0.04 &   21.5 & 30.3 \\
   161-314   &\nodata  &   \nodata &  779 & 834.7  &   93 & $ 0.45\pm$0.44 & $ 0.07\pm$0.17 &   22.1 & 29.1 \\
   163-317   &    K0-K7 &   \nodata &  787 & 834.7  & 1115 & $ 0.35\pm$0.07 & $-0.03\pm$0.04 &   22.4 & 30.2 \\
  d163-222   &   $>=$M2 &   \nodata &  799 & 831.1  &   18 & $-0.80\pm$0.25 &    \nodata   &   21.2 & 27.8 \\
   163-249   &    M1.5e &  2.24 &  800 & 832.9  &   48 & $ 0.15\pm$0.28 & $-0.17\pm$0.25 &   22.4 & 28.9 \\
   164-511   &     M1.5 &  2.26 &  803 & 832.9  & 3535 & $-0.40\pm$0.02 & $-0.15\pm$0.03 &   21.6 & 30.2 \\
   165-235   &       M4 &  1.44 &  807 & 831.1  & 1447 & $-0.56\pm$0.02 & $-0.30\pm$0.05 &   21.4 & 29.7 \\
   166-519   &       M2 &  2.34 &  814 & 832.9  &  207 & $-0.60\pm$0.06 & $-0.57\pm$0.14 &   20.8 & 28.7 \\
   166-316   &\nodata  &   \nodata &  820 & 834.7  &  152 &    \nodata         & $ 0.50\pm$0.10 &   22.9 & 30.1 \\
  d167-231   &       M4 &  0.41 &  825 & 831.1  & 3641 & $-0.43\pm$0.02 & $-0.17\pm$0.03 &   21.6 & 30.2 \\
   167-317   &    G4-K5 &   \nodata &  826 & 834.7  & 1585 & $ 0.05\pm$0.04 & $ 0.07\pm$0.03 &   22.1 & 30.4 \\
   168-328   &\nodata  &   \nodata &  827 & 834.7  &  812 & $-0.18\pm$0.04 & $-0.58\pm$0.06 &   22.2 & 30.5 \\
  d170-249   &    K5-M2 &   \nodata &  844 & 831.1  &  227 & $-0.26\pm$0.08 & $-0.45\pm$0.11 &   22.1 & 29.2 \\
   170-337   &      M2e &   \nodata &  847 & 834.7  &  216 & $-0.61\pm$0.06 & $-0.30\pm$0.14 &   21.6 & 29.2 \\
  d171-340   &      K8e &   \nodata &  856 & 834.7  & 3313 & $-0.43\pm$0.02 & $-0.31\pm$0.03 &   21.7 & 30.3 \\
   171-334   &    K0-K2 &  2.47 &  855 & 834.7  & 3364 & $-0.06\pm$0.02 & $-0.02\pm$0.02 &   21.7 & 31.8 \\
  d172-028   &       M3 &   \nodata &  865 & 811.7  &  335 & $-0.50\pm$0.05 & $-0.44\pm$0.10 &   21.4 & 29.0 \\
  d174-236   &    G4-K5 &   \nodata &  876 & 829.4  &  253 & $ 0.46\pm$0.12 & $ 0.44\pm$0.06 &   22.7 & 29.8 \\
   174-414   &       M5 &  2.39 &  887 & 581.8  &  229 & $-0.51\pm$0.06 & $-0.45\pm$0.12 &   21.8 & 29.3 \\
   175-251   &\nodata  &   \nodata &  884 & 831.1  &  244 & $ 0.57\pm$0.40 & $ 0.79\pm$0.05 &   23.0 & 29.9 \\
  d176-543   &       K8 &  2.48 &  901 & 829.4  & 2303 & $-0.29\pm$0.02 & $-0.12\pm$0.03 &   21.4 & 30.0 \\
   176-325   &\nodata  &   \nodata &  900 & 832.9  &   22 & $-0.20\pm$0.68 & $ 0.61\pm$0.33 &   23.5 & 30.4 \\
   177-454   &    M5.5e &  1.52 &  914 & 390.8  &  412 & $-0.71\pm$0.04 & $-0.43\pm$0.10 &   21.4 & 29.5 \\
  d181-825   &     M1:e &  0.34 &  948 & 795.8  &  487 & $-0.28\pm$0.06 & $ 0.49\pm$0.05 &   20.9 & 30.0 \\
   182-316   &       M2 &  3.75 &  955 & 831.1  &  170 & $-0.24\pm$0.09 & $-0.21\pm$0.12 &   21.5 & 28.8 \\
  d183-405   &       M3 &  2.55 &  966 & 783.4  & 1336 & $-0.73\pm$0.02 & $-0.36\pm$0.06 &   21.7 & 30.6 \\
   184-427   &     M2.5 &  4.03 &  967 & 786.9  &  253 & $-0.70\pm$0.05 & $-0.72\pm$0.15 &   21.4 & 29.0 \\
   189-329   &      M0e &  4.48 & 1000 & 486.3  &  315 & $-0.27\pm$0.06 & $-0.20\pm$0.08 &   22.0 & 29.6 \\
   191-350   &    G8-K5 &  2.46 & 1011 & 530.5  &  354 & $-0.03\pm$0.08 & $ 0.40\pm$0.06 &   22.1 & 29.6 \\
  d197-427   & M0-M2.5e &  0.20 & 1045 & 834.7  & 4273 & $-0.26\pm$0.02 & $-0.11\pm$0.02 &   21.8 & 30.4 \\
   198-222   &   late-M &   \nodata & 1056 & 825.8  &  359 & $ 0.55\pm$0.07 & $ 0.19\pm$0.06 &   22.5 & 29.8 \\
   198-448   &  M1-M6.5 &   \nodata & 1058 & 778.1  & 1348 & $-0.46\pm$0.03 & $-0.37\pm$0.04 &   21.6 & 29.8 \\
   202-228   &\nodata  &   \nodata & 1084 & 825.8  &  651 & $-0.37\pm$0.04 & $-0.34\pm$0.06 &   22.0 & 29.8 \\
  d203-504   &\nodata  &   \nodata & 1091 & 746.2  &   78 & $ 0.19\pm$0.14 & $-0.37\pm$0.14 &   22.6 & 29.5 \\
   205-330   &       M0 &  2.02 & 1101 & 371.4  & 3753 & $-0.41\pm$0.02 & $-0.23\pm$0.03 &   21.7 & 30.6 \\
   205-052   &\nodata  &   \nodata & 1104 & 809.9  & 1907 & $-0.07\pm$0.03 & $-0.17\pm$0.03 &   22.0 & 30.2 \\
  d205-421   &     cont &   \nodata & 1107 & 832.9  &   60 & $-0.23\pm$0.15 & $-0.42\pm$0.23 &   22.0 & 28.7 \\
  d206-446   &      M2e &   \nodata & 1112 & 832.9  & 2788 & $ 0.20\pm$0.02 & $-0.03\pm$0.02 &   22.3 & 30.5 \\
   208-122   &       K7 &  0.87 & 1120 & 815.2  &  323 & $-0.18\pm$0.06 & $-0.16\pm$0.08 &   21.8 & 29.2 \\
   212-557   &     M0.5 &   \nodata & 1139 & 498.7  &  307 & $ 0.61\pm$0.12 & $ 0.70\pm$0.04 &   23.0 & 30.4 \\
   212-260   &       M3 &  3.01 & 1141 & 454.5  &  874 & $-0.25\pm$0.04 & $-0.29\pm$0.05 &   22.1 & 30.4 \\
   213-346   &       K7  & 2.09 & 1149 & 615.4  & 4491 & $-0.43\pm$0.01 & $-0.25\pm$0.02 &   21.7 & 30.5 \\
   215-317   &     M3.5 &  4.74 & 1155 & 514.6  &  348 & $-0.36\pm$0.05 & $-0.58\pm$0.08 &   21.6 & 29.4 \\
   218-339   &    K5-K7 &  3.97 & 1167 & 779.8  &  335 & $ 0.33\pm$0.07 & $ 0.06\pm$0.06 &   22.4 & 29.9 \\
  d218-354   &    G6-K3 &  1.51 & 1174 & 613.6  &  786 & $-0.35\pm$0.04 & $-0.37\pm$0.05 &   21.7 & 29.8 \\
   221-433   &\nodata  &   \nodata & 1184 & 831.1  &  296 & $-0.56\pm$0.05 & $-0.28\pm$0.11 &   21.3 & 29.0 \\
   224-728   &    M4.5e &   \nodata & 1206 & 801.1  &  914 & $-0.39\pm$0.03 & $-0.27\pm$0.05 &   21.8 & 29.8 \\
   236-527   &\nodata  &   \nodata & 1262 & 822.3  &   60 & $ 0.07\pm$0.27 & $ 0.58\pm$0.12 &   22.5 & 28.9 \\
   237-627   &       M3 &  3.21 & 1263 & 270.6  &   29 & $-0.78\pm$0.14 & $-0.91\pm$0.75 &   21.7 & 28.8 \\
   239-510   &       M1 &   \nodata & 1275 & 824.1  &  564 & $ 0.02\pm$0.05 & $-0.07\pm$0.05 &   22.1 & 29.7 \\
   240-314   &\nodata  &   \nodata & 1276 & 790.5  &  578 & $ 0.28\pm$0.06 & $ 0.17\pm$0.05 &   22.3 & 29.9 \\
   242-519   &   K0-K5e &  1.72 & 1281 & 820.5  & 1554 & $-0.58\pm$0.02 & $-0.34\pm$0.05 &   21.3 & 29.7 \\
  d244-440   &      M0e &  0.92 & 1290 & 626.0  & 1930 & $ 0.02\pm$0.03 & $-0.04\pm$0.03 &   22.2 & 30.5 \\
   245-632   &    M4-M5 &  1.39 & 1291 & 328.9  &  406 & $-0.48\pm$0.05 & $-0.38\pm$0.08 &   21.5 & 29.5 \\
   245-502   &      M1e &   \nodata & 1293 & 661.4  &   20 & $-0.13\pm$0.36 & $ 0.18\pm$0.34 &   21.9 & 28.2 \\
   247-436   &      M0e &  2.15 & 1302 & 774.5  &  351 & $-0.11\pm$0.06 & $ 0.02\pm$0.07 &   22.0 & 29.4 \\
   250-439   &\nodata  &   \nodata & 1313 & 774.5  &   94 & $-0.70\pm$0.08 & $-0.60\pm$0.29 &   21.0 & 28.4 \\
  d252-457   &\nodata  &   \nodata & 1317 & 680.8  &   20 & $ 0.21\pm$0.34 & $-0.05\pm$0.32 &   21.4 & 28.1 \\
 d280-1720   &       M4 &  0.64 & 1404 & 751.6  &  618 & $-0.69\pm$0.03 & $-0.65\pm$0.16 &   21.1 & 29.3 \\
   282-458   &   K6-K8e &   \nodata & 1409 & 808.1  & 6381 & $-0.02\pm$0.01 & $-0.06\pm$0.02 &   22.0 & 30.7 \\
\enddata

\tablenotetext{a}{See Getman et al.\ 2005a.}
\tablenotetext{b}{Effective exposure time.}
\tablenotetext{c}{Net photon counts after background subtraction.}
\tablenotetext{d}{COUP X-ray hardness ratio 2, defined as
  $(C_s - C_m)/(C_s + C_m)$ where $C_s$ is counts in the
  0.5--1.7 keV band and  $C_m$ is counts in the
  1.7--2.8 keV band.}
\tablenotetext{e}{COUP X-ray hardness ratio 3, defined as
  $(C_m - C_h)/(C_m + C_h)$ where $C_m$ is counts in the 
  1.7--2.8 keV band and $C_h$ is counts in the
  2.8--8.0 keV band.}
\tablenotetext{f}{Absorbing column derived from spectral model fitting.}
\tablenotetext{g}{Total X-ray luminosity in the 0.5--8.0 keV
  band, corrected for absorption, as derived from spectral
  model fitting.} 

\end{deluxetable}
}

{
\begin{deluxetable}{cccc}
\tabletypesize{\footnotesize}
\tablewidth{0pt}
\tablecaption{Statistics of Proplyd X-ray Counterparts}
\tablehead{
\colhead{Group\tablenotemark{a}}&
\colhead{Total No.}& 
\colhead{$\Delta_X<1.0''$}& 
\colhead{$\Delta_X<0.4''$}
}
\startdata
candidates   &  172         &    112      &    105 \\
rejected     &   22         &      9      &      8 \\
 & & & \\
proplyds     &  143\tablenotemark{b}     &      101     &      94\\
 & & & \\
star   &  106         &     90      &     84\\
no star   &   37         &     11      &     10\\
 & & & \\
jet(s)   &  30\tablenotemark{c} &  19 & 19 \\
 & & & \\
dark disk  & 39\tablenotemark{d}    &        22    &       19\\
dark disk, no star & 21         &      4    &        3\tablenotemark{e}\\
 & & & \\
near-IR src    &119\tablenotemark{f}    &        90     &      ..  \\
no near-IR src & 10         &      1    &        1\tablenotemark{g}\\
\enddata

\tablenotetext{a}{Based on positions and appearances in ACS
  images; ``star'' (``no star'') indicates those proplyds
  with (without) a central star apparent in ACS images.}
\tablenotetext{b}{Not including 6 proplyd candidates (3 with COUP
  counterparts) that lie outside the ACS fields, and not including the
  jet/disk source d347-1535, which lies outside the COUP field.}
\tablenotetext{c}{Not including d347-1535.}
\tablenotetext{d}{Not including two proplyds lying outside the COUP field.}
\tablenotetext{e}{COUP IDs: 419, 476, and possibly 948}
\tablenotetext{f}{IR source within $0.4''$ (except d239-334; IR source found $0.8''$ away); 14
  proplyd candidates not in VLT field} 
\tablenotetext{g}{COUP ID: 695}

\end{deluxetable}
}

{
\begin{deluxetable}{ccccccl}
\tablewidth{0pt}
\tabletypesize{\footnotesize}
\tablecaption{Silhouette Disks Detected in ACS Images}
\tablehead{
\colhead{Object}&
\colhead{Dimensions}& 
\colhead{$R$\tablenotemark{a}}& 
\colhead{P.A.}& 
\colhead{Star?}&
\colhead{$\log{N_H}$}&
\colhead{Comments\tablenotemark{b}} \\
 & ($''$) & & ($^\circ$) & &  (cm$^{-2}$) &
}
\startdata
d053-717  &  1.1$\times$0.2  & 5.5 &  110  &  Y   &   22.70 &  COUP 241; edge-on? (see \S 4.1) \\
d072-135  &  1.0$\times$0.25 & 4.0 &  100  &  N   &   ...   &  edge-on?           \\
d109-327  &  0.2$\times$0.1  & 2.0 &  160  &  N   &   ...   &             \\
d114-426  &  2.7$\times$0.7  & 3.9 &  30   &  N   &   23.73 &  COUP 419; edge-on? [B2000: i $>$ 85 deg] \\
d121-1925 &  0.8$\times$0.5  & 1.6 &  120  &  Y   &   21.57 &  COUP 460 [B2000: i = 51 deg] \\
d124-132  &  0.3$\times$0.1  & 3.0 &  0    &  N   &   ...   &  COUP 476; poor spectral fit \\
131-046   &  0.3$\times$0.2  & 1.5 &  80:  &  N   &   ...   &         \\
d132-042  &  0.4$\times$0.25 & 1.6 &  85   &  N   &   ...   &          \\
d132-1832 &  1.5$\times$0.3  & 5.0 &  60   &  N   &   ...   &  edge-on? [B2000: i = 75 deg] \\
140-1952  &  0.5$\times$0.5  & 1.0 &  ..   &  Y   &   20.00 &  COUP 597 \\
d141-520  &  0.4$\times$0.35 & 1.1 &  135  &  Y   &   21.16 &  COUP 604 \\
d143-522  &  0.4$\times$0.2  & 2.0 &  140  &  N   &   ...   &         \\
d147-323  & 0.25$\times$0.15 & 1.7 &  40   &  Y   &   22.23 &  COUP 658 \\
d154-240  & 0.25$\times$0.10 & 2.5 &  90   &  N   &   ...   &           \\
d163-026  &  0.5$\times$0.15 & 3.3 &  160  &  N   &   ...   &  edge-on? [B2000: i $>$ 78 deg]  \\
d163-222  &  0.3$\times$0.2  & 1.5 &  70   &  Y   &   21.20 &  COUP 799 \\
d165-254  &  0.4$\times$0.2  & 2.0 &  5    &  N   &   ...   &  [B2000: i $>$ 71 deg] \\
166-519   &   ??             &  ?? &  ??   &  N?  &   ...   &  orientation, dimensions uncertain \\
d167-231  &  0.4$\times$0.4  & 1.0 &  ..   &  Y   &   21.62 &  COUP 825 [B2000: i $<$ 30 deg] \\
d172-028  &  0.6$\times$0.4  & 1.5 &  5    &  Y   &   21.35 &  COUP 865 [B2000: i = 55 deg] \\
d176-543  &  0.6$\times$0.3  & 2.0 &  20   &  Y   &   21.43 &  COUP 901 \\
d177-541  &   ??             & ??  &  ??   &  N   &   ...   &  orientation, dimensions uncertain \\
d181-247  &  0.3$\times$0.15 & 2.0 & 160  &  N   &   ...   &         \\
d181-825  &  1.5$\times$0.6  & 2.5 & 70   &  N?  &   22.78\tablenotemark{c} &  COUP 948 \\
d182-332  &  0.3$\times$0.15 & 2.0 & 0    &  Y?  &   ...   &  [B2000: i = 60 deg] \\
d182-413  &  0.5$\times$0.15 & 3.3 & 90   &  N   &   ...   &         \\
d183-419  &  0.3$\times$0.15 & 2.0 & 40   &  N   &   ...   &          \\
d183-405  &  0.7$\times$0.5  & 1.4 & 45   &  Y   &   21.66 &  COUP 966 [B2000: i = 39 deg] \\
d191-232  &  0.3$\times$0.1  & 3.0 & 170  &  N   &   ...   &  [B2000: i = 65 deg] \\
d197-427  &  0.6$\times$0.4  & 1.5 & 50   &  Y   &   21.80 &  COUP 1045 \\
202-228   &  0.2$\times$0.15 & 1.3 & 45   &  Y   &   21.97 &  COUP 1084 \\
d203-506  &  0.4$\times$0.2  & 2.0 & 15   &  N   &   ...   &  [B2000: i = 67 deg] \\
d205-421  &  0.4$\times$0.3  & 1.3 & 60   &  Y   &   22.02 &  COUP 1107 \\
d206-446  &  0.5$\times$0.3  & 1.7 & 70   &  Y   &   22.30 &  COUP 1112 \\
d218-354  &  1.4$\times$0.6  & 2.5 & 70   &  Y   &   21.67 &  COUP 1174 [B2000: i = 65 deg] \\
d218-529  &  0.4$\times$0.2  & 2.0 & 175  &  N   &   ...   &  [B2000: i = 60 deg] \\
d239-334  &  0.5$\times$0.2  & 2.5 & 20   &  ..  &   ...   &  [B2000: i = 66 deg] \\
d253-1536 &  1.2$\times$0.6  & 2.0 & 80   &  Y?  &   ...   &  \\
d280-1720 &  0.7$\times$0.6  & 1.2 & 10   &  Y   &   21.10 &  COUP 1404 \\
d294-606  &  1.0$\times$0.25 & 4.0 & 85   &  N   &   ...   &  edge-on? [B2000: i $>$ 85 deg] \\
d347-1535 &  0.7$\times$0.2  & 3.5 & 130  &  N   &   ...   &  off COUP FOV \\
\enddata

\tablenotetext{a}{Ratio of major to minor axes of silhouette
disk.}
\tablenotetext{b}{[B2000]: included in Table 1 of Bally et al. 2000}
\tablenotetext{c}{Absorbing column inferred for hard
spectral component.}

\end{deluxetable}
}

{\footnotesize
\begin{deluxetable}{cccl}
\tabletypesize{\footnotesize}
\tablewidth{0pt}
\tablecaption{Jets and Microjets Detected in ACS and/or WFPC2 Images}
\tablehead{
\colhead{Object}&
\colhead{COUP ID}& 
\colhead{Star?}&
\colhead{Comments\tablenotemark{a}} \\
}
\startdata
069-600  &  279 & Y & w069-600 [B2000] \\
d109-327 &  ... & N & HH 510 [B2000] \\
d110-3035&  ... & N & \\
d124-132 &  476 & N & [S2005]\\
131-247  &  524 & N & HH 511 [B2000] \\
d132-042 &  ... & N?& dark disk [S2005]\\
154-324  &  ... & Y & star + jet \\
157-533  &  728 & Y & HH 512 [B2000] \\
164-511  &  803 & Y & jet?\\
165-235  &  807 & Y & HH 513 [B2000] \\
167-317  &  826 & Y & HH candidate [B2000] \\
170-337  &  847 & Y & HH 514 [B2000] \\
d176-543 &  901 & Y & dark disk; HH 515 [B2000] \\
d177-341 & ...  & Y & HH candidate \\
d181-825 &  948 & N?& dark disk; Beehive Proplyd; HH 540\\
d182-413 &  ... & N & dark disk; HH 517 [B2000] \\
191-350  & 1011 & Y & HH candidate [B2000] \\
201-534  & ...  & Y & jet? \\
d203-504 & 1091 & Y & HH 519 \\
d203-506 &  ... & N & dark disk; HH 520 [B2000] \\
d206-446 & 1112 & Y & dark disk; HH 521 [B2000] \\
208-122  & 1120 & Y & jet? \\
d218-354 & 1174 & Y & dark disk; HH candidate [B2000] \\
d218-529 &  ... & N & \\
d239-334 &  ... & ..& HH 522; COUP 1268 at companion [B2000] \\
d244-440 & 1290 & Y & HH 524 [B2000] \\
247-436  & 1302 & Y & HH 525 [B2000] \\
d252-457 & 1317 & Y & HH 526 [B2000] \\
d253-1536&  ... & Y?& HH 668 [S2005]; COUP 1316 at companion \\
282-458  & 1409 & ..& HH 527 [B2000] \\
d347-1535&      & N & dark disk, extensive bipolar jets [S2005]\\
\enddata

\tablenotetext{a}{[B2000]: listed in Table 3 of Bally et al.\ 2000; [S2005]: listed in Smith et al.\ 2005}

\end{deluxetable}
}


\begin{figure}
\includegraphics[scale=0.8,angle=0]{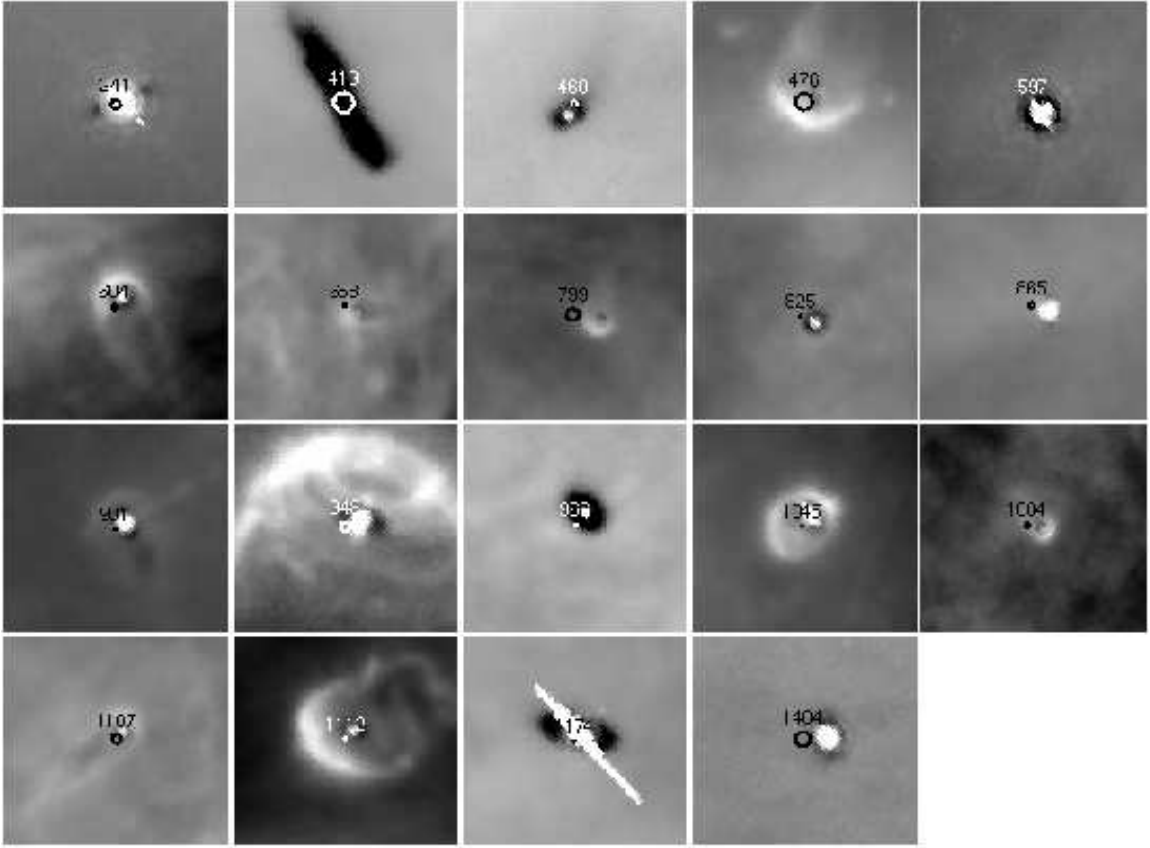} 
\caption{HST/ACS images of silhouette disk proplyds with
  COUP X-ray counterparts. Top row, from left to
  right: d218-354 (COUP 241), d114-426 (419), d121-1925
  (460), 140-1952 (597), d141-520 (604); second row: d147-323 (658),
  d163-222 (799), d167-231 (825), d172-028 (865), d176-543
  (901); third row: d181-825 (948), d183-405 (966), d197-427 (1045),
  202-228 (1084), d205-427 (1045); bottom row: d205-421 (1107), d206-446
  (1112), d218-354 (1174), d280-1720 (1404).  
  The field of view in each displayed ACS 
  image region is $2.7''\times2.7''$ with N up and E to the
  left. In each image, the COUP 
  source position is indicated by a circle whose radius is equal
  to the positional uncertainty for that source, as determined from
  X-ray source detection; there is an additional random
  scatter of $\sim0.2''$ in COUP source positions (see Getman et
  al.\ 2005a, their Fig.\ 9). }  
\label{fig:darkdisk_imgs}
\end{figure}

\begin{figure}
\includegraphics[scale=1.,angle=90]{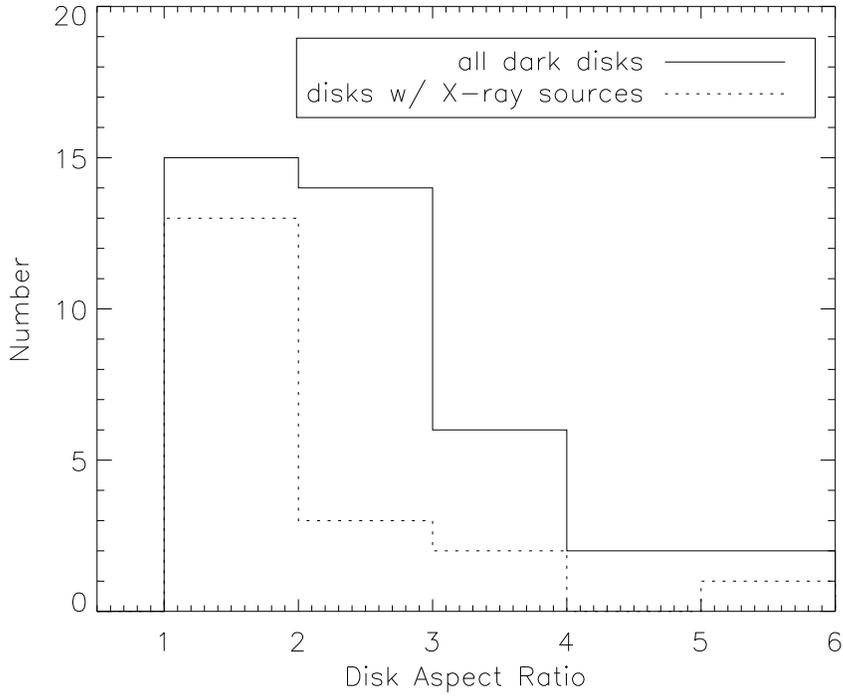} 
\caption{Histograms of number of silhouette disks vs.\
  aspect ratio, for silhouette proplyds listed in Table
  5. The solid line indicates the total number of disks in
  each aspect ratio bin, while the dotted line indicates only those disks
  harboring X-ray sources. } 
\label{fig:histaspect}
\end{figure}


\begin{figure}
\includegraphics[scale=.75,angle=0]{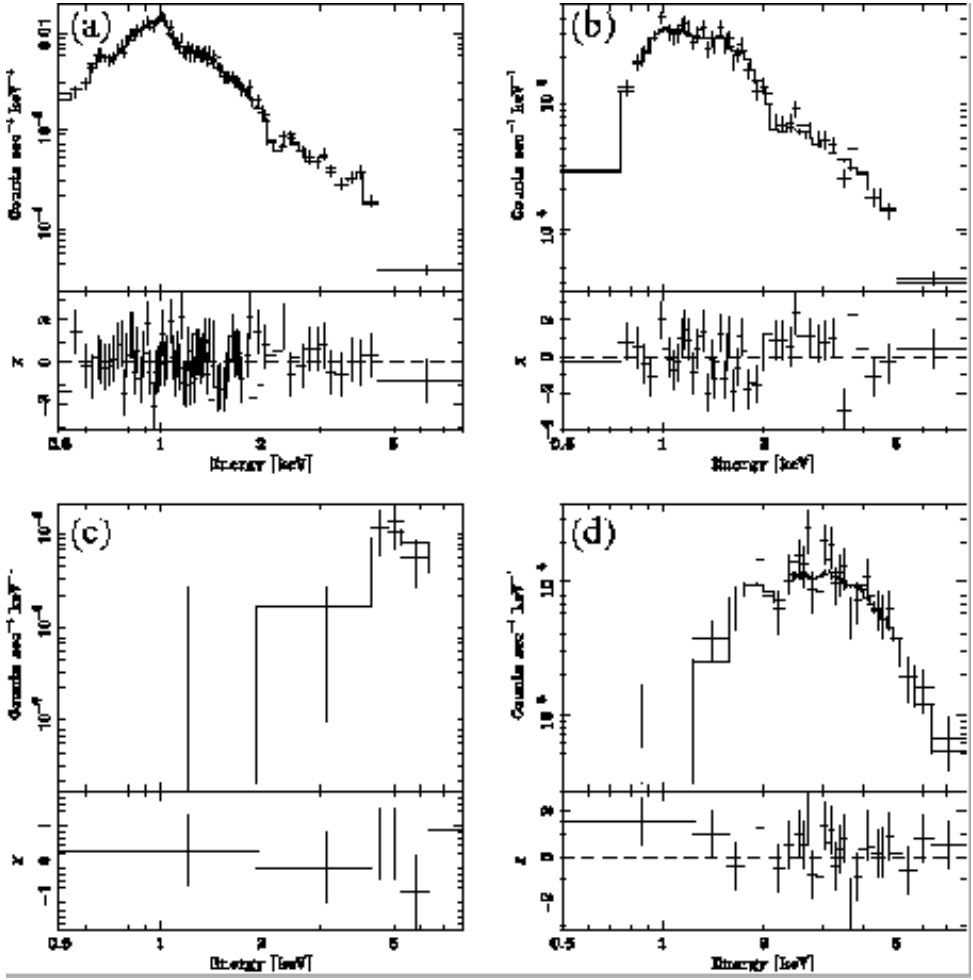} 
\caption{
  COUP X-ray spectra of representative sources associated with 
  silhouette disk proplyds. COUP 597 (panel a) and
  COUP 825 (panel b) are X-ray
  counterparts to presumably nearly face-on disks, i.e.,
  silhouette disks with small aspect 
  ratios, while COUP 419 (panel c) and COUP 241 (panel d) are counterparts
  to silhouette disks with the largest aspect ratios. For
  each source, X-ray spectral data (and associated
  uncertainties) are indicated 
  by crosses, the histogram represents the best-fit
  spectral model, and residuals of the fit are indicated in the
  bottom panel (see Getman et al.\ 2005a). } 
\label{fig:darkdisk_spec}
\end{figure}

\begin{figure}
\includegraphics[scale=1.,angle=90]{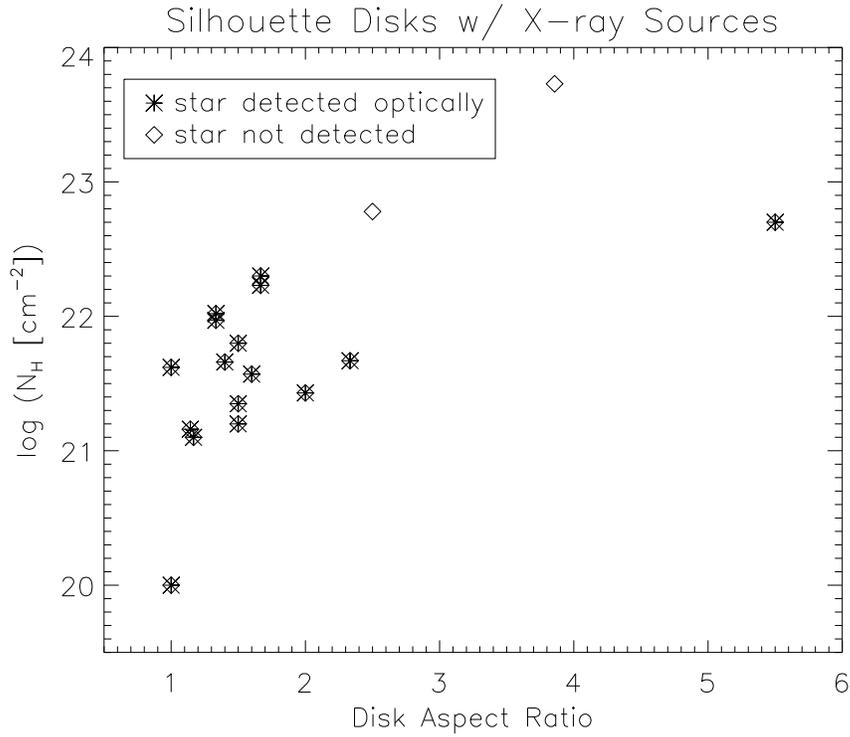} 
\caption{X-ray-inferred column density, $\log{N_{\rm H}}$ (cm$^{-2}$),
  vs.\ disk
  aspect ratio, for silhouette disk proplyds listed in Table
  5. Asterisks indicate proplyds with 
  optically detected central stars. The bright central star
  detected within the silhouette disk with
  the largest aspect ratio, d053-717, may not be the source
  of X-ray emission (see \S 4.1).} 
\label{fig:NHaspect}
\end{figure}

\begin{figure}
\includegraphics[scale=.8,angle=0]{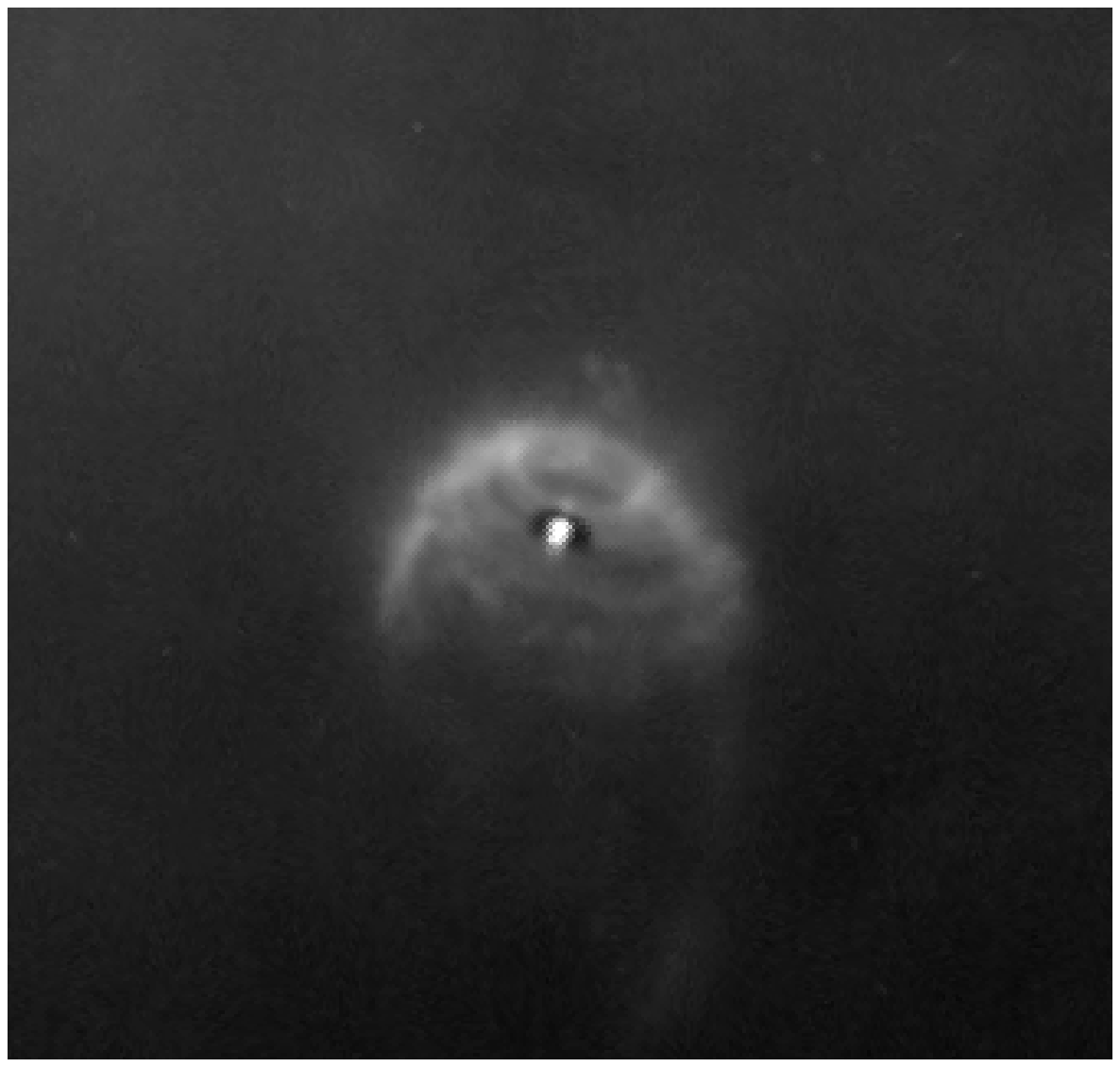} 
\includegraphics[scale=.5,angle=0]{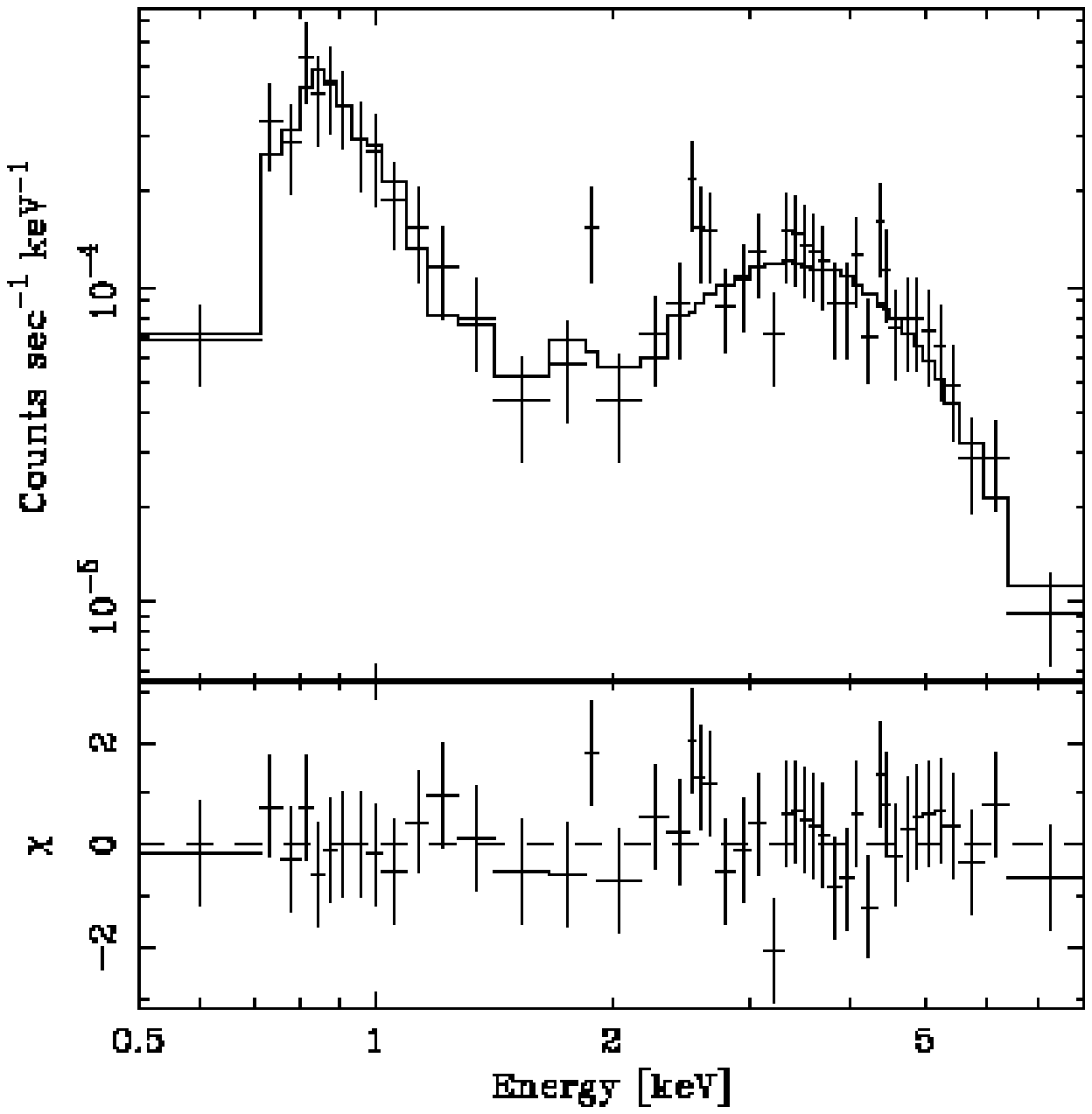} 
\caption{Left: HST/ACS image of the Beehive Proplyd
  (d181-825), associated with COUP 948. The field of view
  $11''\times11''$ with N up and E to the left. 
  Right: COUP X-ray spectrum of COUP 948. Note the double-peaked X-ray
  spectral energy distribution.} 
\label{fig:Beehive}
\end{figure}

\begin{figure}
\includegraphics[scale=.8,angle=0]{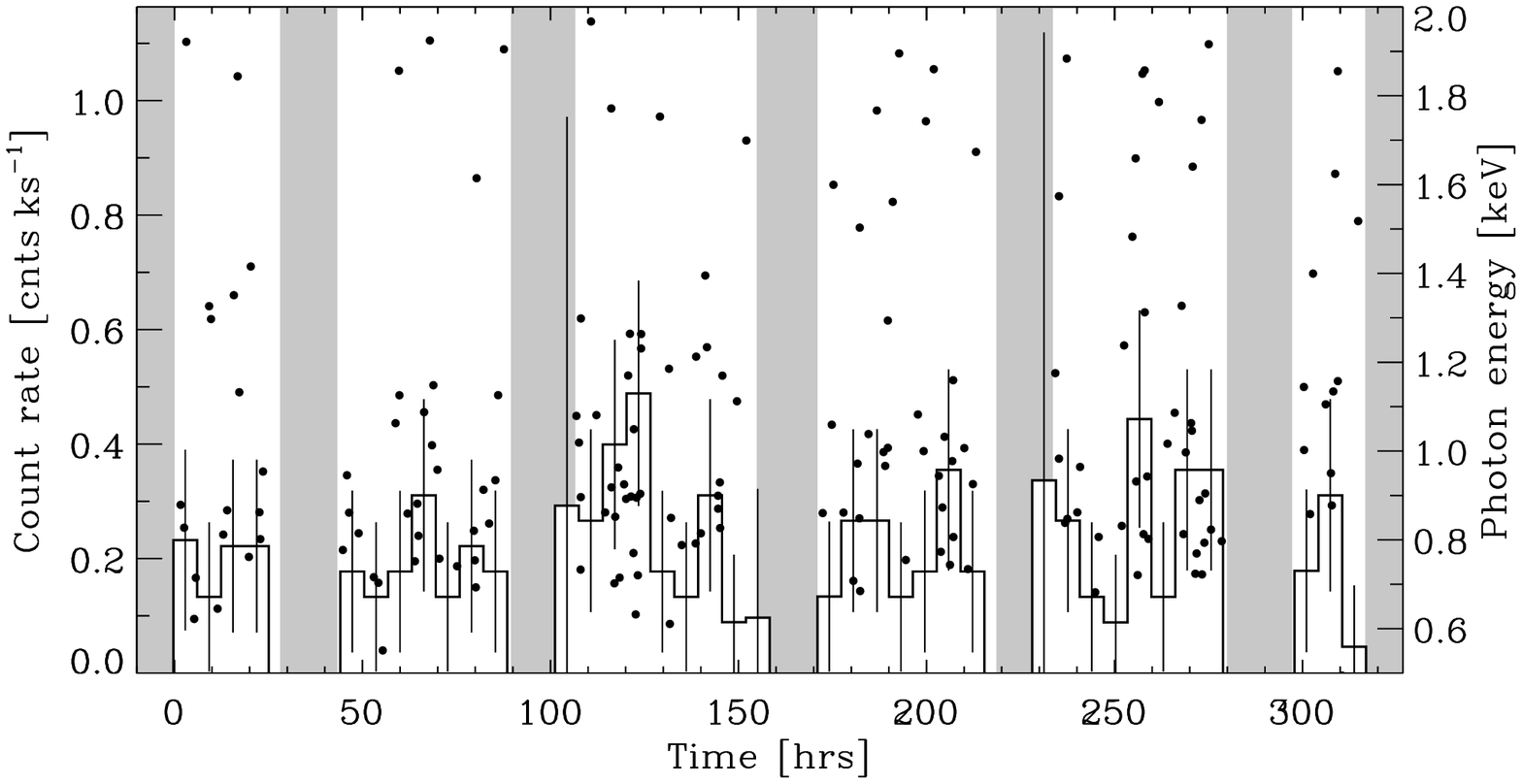} 
\includegraphics[scale=.8,angle=0]{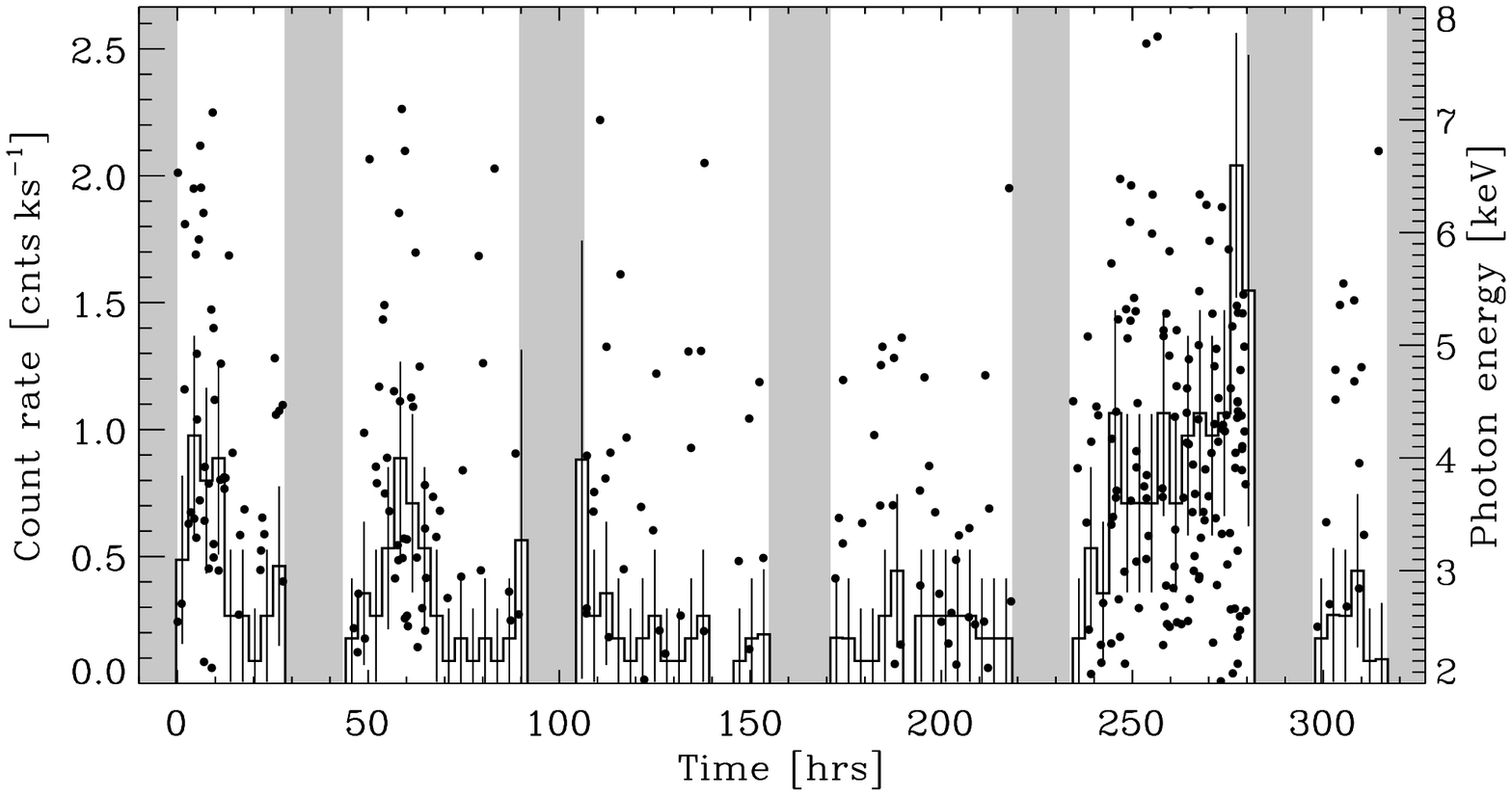} 
\caption{Light curves of COUP 948, which is associated with
  the Beehive Proplyd (d181-825). Top: soft band (0.5--2.0
  keV). Bottom: hard band (2.0--8.0 keV). In each plot, the
  histogram indicates the integrated counts within the band,
  while the points indicate the arrival times and energies of
  individual photons. The grey bands indicate gaps in
  temporal coverage during the 838 ks COUP observation.} 
\label{fig:Beehive_lc}
\end{figure}

\begin{figure}
\includegraphics[scale=.8,angle=90]{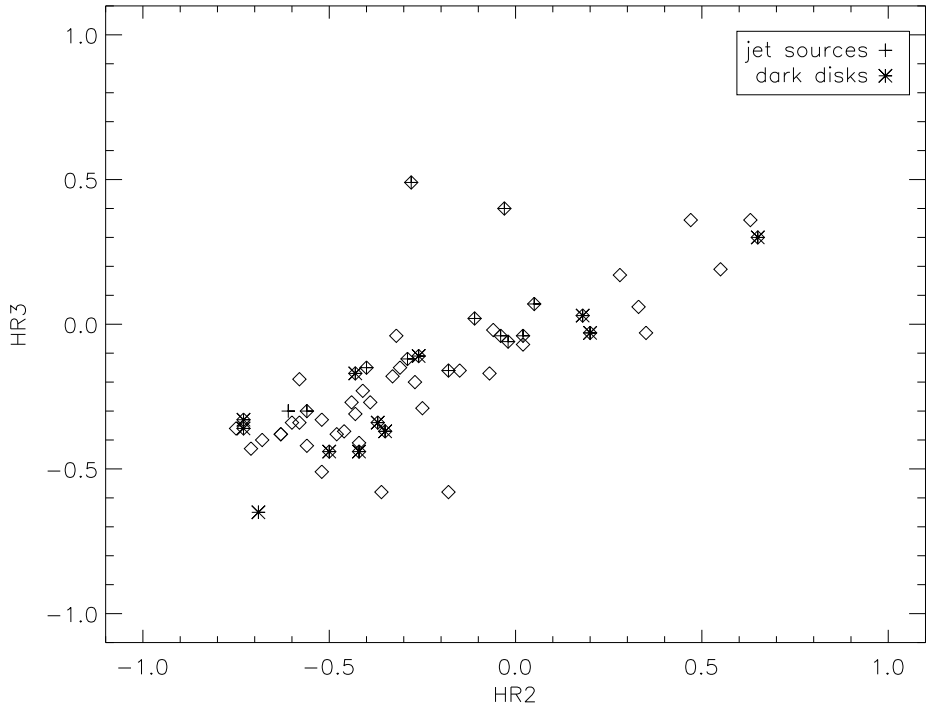} 
\caption{COUP X-ray hardness ratios HR2 and HR3 (see
  Table 3) of proplyd COUP 
  sources (diamonds), with hardness ratios
  of silhouette disk proplyds (``dark disks'') and jet
  sources indicated by asterisks and crosses,
  respectively. Only sources with uncertainties $\le 0.1$ in
  HR2 and HR3 are plotted. The two crosses at the upper left of the
  diagonal locus of points are the COUP counterparts to jet
  sources d181-825 (COUP 948, which has the largest value of
  HR3 of any source in the Figure) and 191-350 (COUP 1011).}  
\label{fig:jetdisk_HRs}
\end{figure}

\begin{figure}
\includegraphics[scale=.6,angle=0]{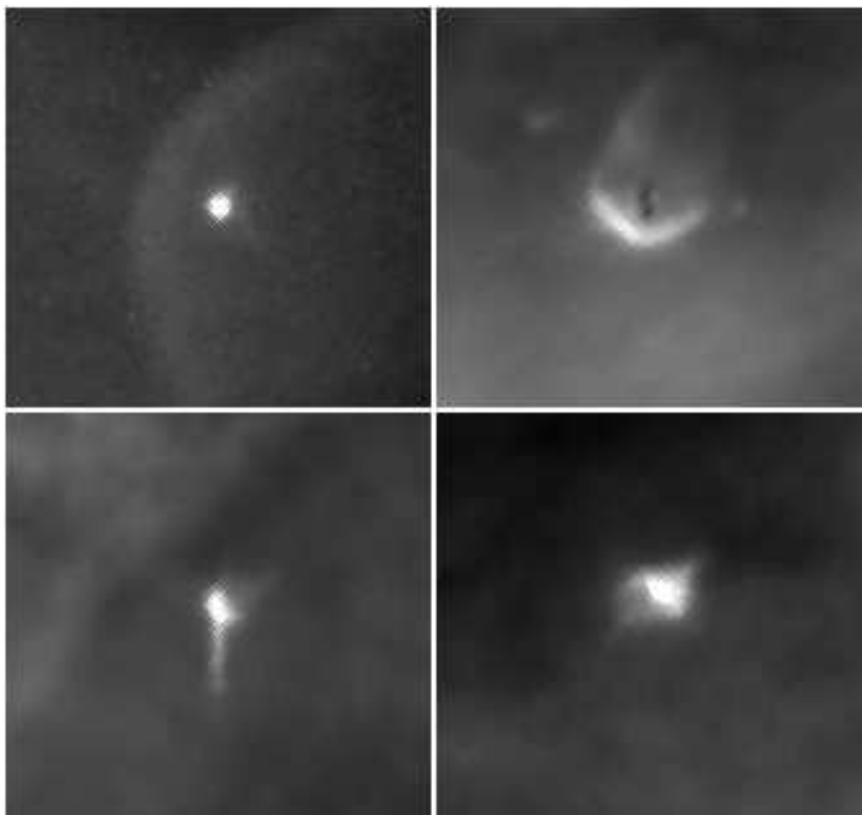} 
\caption{HST/ACS images of jet
  sources whose X-ray hardness ratios appear anomalous. From
  top to bottom, left to right: 069-601,
  d124-132, 131-247, and 191-350 
  (COUP X-ray counterparts are 279, 476, 524, and 1011,
  respectively). The field of view in each case is
  $4.2''\times4.2''$, with N up and E to the left.} 
\label{fig:other_jets}
\end{figure}

\begin{figure}
\includegraphics[scale=.9,angle=0]{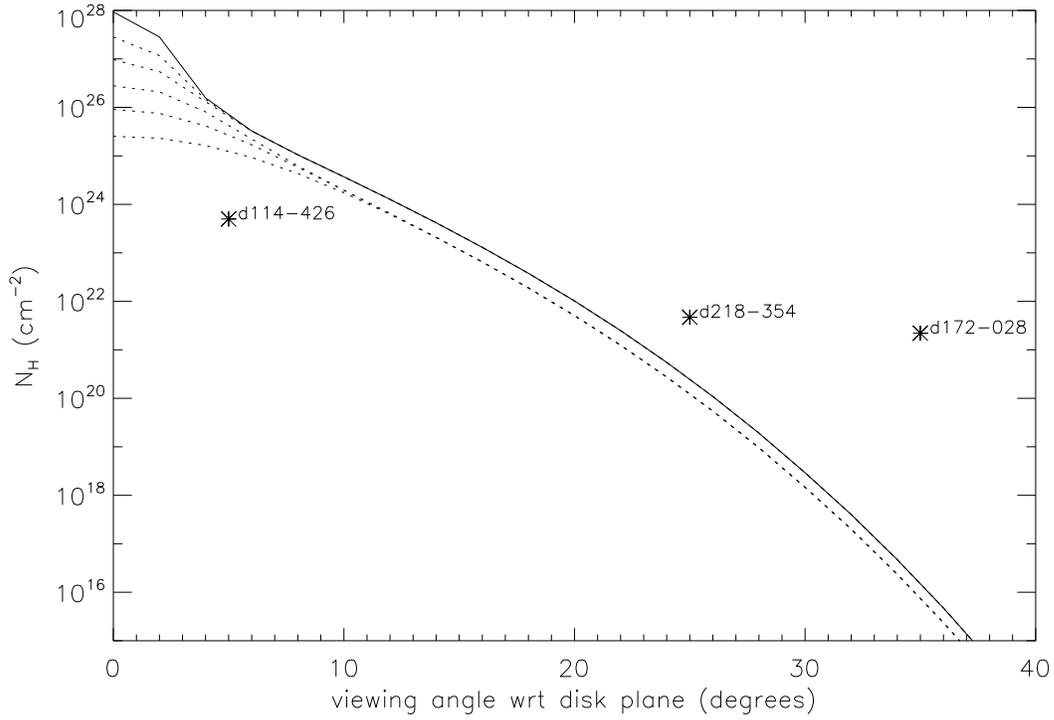} 
\caption{Model distribution of $N_{\rm H}$ vs.\ $i$ obtained from
the circumstellar disk model. The
top-to-bottom curves represent a range of assumed inner radii for the
disk, from 0.03 AU to 10 AU. The estimated inclinations and
corresponding $N_{\rm H}$ values for
three representative proplyds are indicated in the plot;
the disk viewing angles for these sources are based on inclination estimates
listed in Bally et al.\ (2000).}  
\label{fig:modelNHvsi}
\end{figure}

\end{document}